\documentclass[
aip,
 amsmath,amssymb,
 reprint,%
]{revtex4-1}

\usepackage{graphicx}
\usepackage{dcolumn}
\usepackage{bm}
\usepackage{enumerate}

\usepackage{setspace,color,url}

\begin{document}

\title{Carbon membranes for efficient water-ethanol separation}
\date{\today}

\author{Simon Gravelle}
\affiliation{Institut Lumi\`ere Mati\`ere, UMR5306 Universit\'e Lyon 1-CNRS, Universit\'e de Lyon 69622 Villeurbanne, France}
\author{Hiroaki Yoshida}
\affiliation{Laboratoire de Physique Statistique, UMR CNRS 8550, Ecole Normale Sup\'erieure, PSL Research University, 
24 rue Lhomond, 75005 Paris, France}
\affiliation{Toyota Central R\&D Labs., Inc., Nagakute, Aichi 480-1192, Japan}
\author{Laurent Joly}
\affiliation{Institut Lumi\`ere Mati\`ere, UMR5306 Universit\'e Lyon 1-CNRS, Universit\'e de Lyon 69622 Villeurbanne, France}
\author{Christophe Ybert}
\affiliation{Institut Lumi\`ere Mati\`ere, UMR5306 Universit\'e Lyon 1-CNRS, Universit\'e de Lyon 69622 Villeurbanne, France}
\author{Lyd\'eric Bocquet}
\email{lyderic.bocquet@ens.fr}
\affiliation{Laboratoire de Physique Statistique, UMR CNRS 8550, Ecole Normale Sup\'erieure, PSL Research University, 
24 rue Lhomond, 75005 Paris, France}

\begin{abstract} 
We demonstrate, on the basis of molecular dynamics simulations, the possibility of an efficient water-ethanol separation
using nanoporous carbon membranes, namely carbon nanotube 
 membranes, nanoporous graphene sheets, and multilayer graphene membranes.
While these carbon membranes are in general permeable to both pure liquids,
they exhibit a counter-intuitive {``self-semi-permeability''} to water in the presence of water-ethanol mixtures.
This originates in a preferred ethanol adsorption in  nanoconfinement
 that prevents water molecules from entering the carbon nanopores.
 An osmotic pressure is accordingly expressed across the carbon membranes for the water-ethanol mixture,
 which agrees with the classic
van't Hoff
type expression.
 This suggests a robust and versatile membrane-based separation,
 built on
a pressure-driven 
reverse-osmosis 
process across these carbon-based membranes.
 In particular, the recent development of large-scale `graphene-oxide' like membranes 
then 
opens an avenue for a versatile and efficient ethanol dehydration using this separation process, 
with possible application for bio-ethanol fabrication.
\end{abstract}

\maketitle

\section{Introduction} 

Ethanol is the most commonly used commercial bio-fuel, promising 
environmental and economic benefits, such as a reduction of consumption
of crude oil and related environmental pollution.~\cite{BB2009,HGG+2006}
However, the fabrication of bio-ethanol requires an unavoidable step: ethanol dehydration.~\cite{HRT+2008}
The separation of ethanol from water is commonly performed by heating
processes, e.g. pervaporation or distillation, and this step
represents the bulk of the cost for the production of ethanol from
biomass.~\cite{MHH+1983}
A reduction of the cost of ethanol dehydration
is thus critical regarding the global production of bio-ethanol which
reached  46 billion liters in 2007 and could grow up to 125
billion liters by 2020.~\cite{BB2009}

Numerous solutions exist for the dehydration of ethanol, for instance, ordinary
distillation, azeotropic distillation, extractive distillation (with
liquid solvent, or with dissolved salt), liquid-liquid
extraction-fermentation hybrid, adsorption and membrane separation.~\cite{HRT+2008}
Currently, pervaporation, which consists in the partial vaporization of
the liquid through a membrane, is considered as one of the most
effective and energy-saving process for the separation of ethanol and
water.~\cite{SYK+1994, MFS1985, NYN1998}
However, this method requires to heat the system up to $\sim 80^\circ$C for water-ethanol separation, 
and, just as every thermal separation method, suffers from the
disadvantage of a high energy penalty, 
associated with
heat losses to the environment, heat losses due to minimal driving
forces, and losses due to boiling point elevation.~\cite{Semiat2008}

On the other hand, membrane-based separation methods, such as
 ultra-filtration and  reverse osmosis (RO), have gained considerable importance,
because they offer superior treatment at modest cost,
high stability and efficiency, and low energy
requirement.~\cite{SGA+2001, IGS+2009}
Particularly, RO is currently the most important desalination
technology~\cite{LAM2011} thanks to a very low cost in comparison with
thermal desalination technology.~\cite{FLW+2007}
The membrane separation is thus seen as a viable and effective technology at both
laboratory and industrial scales. 
However, the applicability of the membrane separation is not obvious when it
comes to two species that are 
neutral and have very similar size, such as ethanol and water.

In this paper, 
we demonstrate, using molecular dynamics (MD) simulations, that 
water-ethanol separation can be achieved with carbon-based membranes.
We show that in the presence of water-ethanol mixtures, nanoporous carbon membranes may 
become fully impermeable to water while 
keeping a high permeability to ethanol. 
This is in spite of these carbon membranes being in general permeable to
both water and ethanol when they are used as pure
components.\cite{Note1} 
In the following, we coin accordingly this behavior ``self-semi-permeability'' to 
highlight the change of the membrane permeability in the presence of mixtures,
which occurs without any further external action.
This counter-intuitive result is highlighted by the existence of an
osmotic pressure for the ethanol-water mixture across the membrane,
which has to be bypassed in order to separate the ethanol from water. 
The basic mechanism for this specific separation lies in a 
preferred adsorption of ethanol as compared to water.
We found a similar separation property
with three different types of carbon
based membranes:  namely, carbon nanotubes; a single graphene sheet
pierced with nanopores; and a multilayer graphene membrane, mimicking the porous
structure of reduced graphene-oxide (GO) membranes.
This highlights the robustness and versatility of the underlying mechanism.

Our results suggest an efficient
membrane-based method for the separation of water from ethanol.
Thanks to the recent progress made for the development of GO membranes,
we believe that this versatile method may offer a new solution for
ethanol dehydration, with a significant potential
impact on the production of bio-ethanol. 


\section{Ethanol-water mixtures across carbon membranes: MD simulations}

\begin{figure}
\begin{center}
\includegraphics[width=\linewidth]{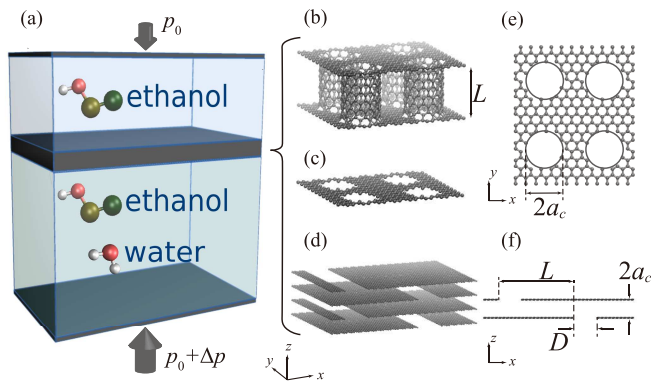}
\caption{
(a) Schematic of the system.
(b) Carbon nanotube membrane. 
(c) Membrane of single pierced graphene sheet.
(d) Multilayer graphene membrane.
(e) Top view of panel (b). 
(f) Side view of panel (d). 
}
 \label{fig:geometry}
\end{center}
\end{figure}

We investigate the hydrodynamic permeability of different carbon-based
membranes to ethanol-water mixtures using  MD
simulations, employing the open source code LAMMPS.~\cite{Plimpton1995,LAMMPS}
The system consists of two reservoirs separated by a carbon-based
membrane, see Fig.~\ref{fig:geometry}(a). 
The top reservoir initially contains pure ethanol while the bottom one
is filled with a mixture of water and ethanol. 
Periodic boundary conditions are imposed in all directions, and two
graphene sheets at the top and bottom ends of the reservoir are used as
a piston to control the pressure in each reservoir.

We consider three membranes with different porous structures:
First, a carbon nanotube (CNT) membrane, consisting of 
two pierced graphene sheets connected by short CNTs 
(length $L = 13$\,{\AA} and radius $a_c$ varied between $3.5$ and
$6.2$\,{\AA}, see Fig.~\ref{fig:geometry}(b) and (e)).
We also modeled a nanoporous graphene membrane, using a single graphene
sheet pierced with nanometric circular pores with radius $a_c$ varied
between $3.5$ and $6.2$\,{\AA} (Fig.~\ref{fig:geometry}(c)). 
Finally, we consider a multilayer graphene membrane, made of 
stacking graphene sheets, pierced with nanoslits of width 
$D=14.1$\,{\AA}
(Fig.~\ref{fig:geometry}(d) and (f)).
The nanoslits are arranged in a staggered fashion with offset 
$L=34.1$\,{\AA},
forming highly ordered
films with 2D nanochannels between the sheets.
The inter-layer distance of this membrane, denoted by $h_c$,
is varied from $6.8$ to $20$\,{\AA}.
This porous structure, consisting of a `millefeuille' of pure graphene sheets, is considered as a simplified 
model for GO membranes. This corresponds merely to `reduced' GO membranes, for which 
the chemical groups covering the graphene sheets can be eliminated.

For the interaction potentials, we employ the TIP4P/2005 water
model.~\cite{AV2005}
The ethanol
molecule is described with the united atom model optimized potentials
for liquid simulations (OPLS).~\cite{JMS1984,Jorgensen1986}
The parameters for the carbon atoms of the wall are extracted from the
AMBER96 force field,~\cite{CCB+1995}
and the Lorentz--Berthelot mixing rules are used 
to determine the Lennard-Jones parameters for the cross-interactions.
Finally, the positions of carbon atoms in the membrane are fixed
and the graphene pistons move as a rigid body. 
Note that simulations with flexible and fixed walls 
have been
shown to give similar
results for the statics and friction of confined
liquids.~\cite{AK2008,TM2009,WWJ+2003}

During simulation runs, the system is maintained at $300$\,K using two
Berendsen thermostats, one in each reservoir.~\cite{BPG+1984}
Those thermostats are applied to molecules at more than $5$\,{\AA} from
the membrane,
so that the flow in the membrane and at the membrane
entrances is not affected by the thermostating procedure.~\cite{TC2015}
The pressures of top and bottom reservoirs are maintained 
at $p_0 = 1$\,bar and $p_0 + \Delta p$, respectively.
After the equilibration for at least $0.1$\,ns
with a plug preventing the exchange of molecules across the membrane,
the time evolution of the number of molecules in the reservoirs is recorded.

\section{Results}

%
\begin{figure*}[t]
 \begin{center}
\includegraphics[width=0.9\linewidth]{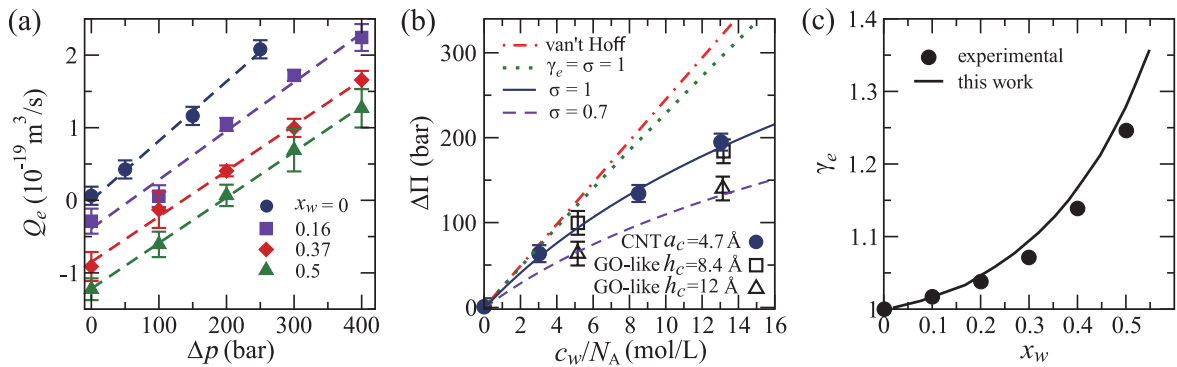}
\caption{
(a) Flow rate of ethanol $Q_e$ per tube in a CNT membrane with
 $a_c=4.7$\,{\AA} as a
 function of the applied pressure difference $\Delta p$ for various
 initial molar fraction of water $x_w$. 
The dashed lines are linear fit, whose slope is the hydrodynamic permeance.
(b) Osmotic pressure $\Delta \Pi$ as a function of the initial
 concentration of water $c_w$ for a CNT membrane and GO-like membranes.
The prediction of Eq.~\eqref{eq:pi} with fitted values
for the activity coefficient $\gamma_e$ is shown by the solid line
 ($\sigma=1$)
and the dashed line ($\sigma=0.7$),
and that with assuming $\gamma_e=\sigma=1$ is shown by the dotted line.
The dash-dotted line indicates the linearized van't Hoff law.
(c) Activity coefficient $\gamma_e$ used in Eq.~\eqref{eq:pi},
for the solid and dashed lines in panel (b), in comparison with
the values taken from Ref.~\citenum{GP2007}.
}
	\label{fig:cnt_flux}
	\end{center}
\end{figure*}

As quoted above, we have explored filtration across three types of carbon membranes: CNT membranes,
graphene membranes pierced with nanopores, and multilayer carbon membranes, mimicking GO membranes.~\cite{YB2016}
In the following we start by investigating CNT membranes and then the results
are generalized to the two other types of carbon membranes. We anticipate that similar results are obtained
for the various types of membranes.

\subsection{Flux and osmotic pressure: CNT membranes}
Under a pressure drop $\Delta p$, the fluxes of ethanol
$Q_{e}$ and 
water $Q_{w}$ are deduced from the linear
fit of the time dependent variation of the number of molecules crossing
the membrane $\Delta N^B_{e,w} (t)$:
\begin{equation}
Q_{e,w} = \dfrac{M_{e,w}}{\rho_{e,w} N_{\rm A}} \dfrac{\mathrm d \Delta
 N^B_{e,w}}{\mathrm d t},
\label{eq:flux}
\end{equation} 
where $M_{e,w}$ and $\rho_{e,w}$ are respectively the molar mass and
density of ethanol and water, and $N_{\rm A}$ is the Avogadro constant.
The flux $Q_e$ of ethanol for a membrane of nanotubes of radius 
$a_c = 4.7$\,{\AA} and length $L=13$\,{\AA} is reported in
Fig.~\ref{fig:cnt_flux}(a),
for varying applied pressure difference $\Delta p$.
The results are plotted for various values of the initial molar fraction of water $x_w$, 
defined as $x_w=N_{w}^B/(N_w^B+N_e^B)$, with $N_{e,w}^B$ being the
initial number of molecules in the bottom reservoir.

\subsubsection{Permeability}
A first feature of Fig.~\ref{fig:cnt_flux}(a) is that 
the flux of ethanol is found to be linear in the applied pressure, 
regardless of the concentration in water.
From the plot, the hydrodynamic permeance $\mathcal{L}$ of pure ethanol ($x_w=0$) is
extracted, which is defined as the flux per unit area normalized 
by the pressure drop: 
\begin{equation}
\mathcal{L} = \dfrac{Q_e}{A \Delta p} \approx 1104\,\text{L}/(\text{m}^2\cdot\text{h}\cdot\text{bar}), 
\end{equation}
where $A$ is the area of the membrane.

This result can be compared to hydrodynamic predictions. 
Since the channel length is relatively short ($L/a_c \sim 1$) and the
slip length of ethanol inside CNTs
is large,~\cite{FSJ+2012} the viscous entrance effect~\cite{SL2011}
is expected to dominate the overall dissipation.
This effect, which originates in the bending of the streamlines toward
the pore, was first discussed by Sampson who calculated the velocity
profile flowing through an infinitely thin membrane pierced with
circular hole.~\cite{Sampson1891}
In this case, the total flow rate $Q$ is linked to the pressure drop
$\Delta p$ through:
\begin{equation}
Q = \dfrac{a^3}{C \eta} \Delta p,
\label{eq:Samp}
\end{equation}
where $a$ is the pore radius (effective radius, see below) and $\eta$ the fluid viscosity. 
$C$ is a numerical constant, which is $C=3$ for no-slip boundary conditions,
but may differ for slipping nanotube surfaces.~\cite{GJY+2014} Under
the present conditions,  $C \approx 1.4$ for
a CNT with radius $a_c = 4.7$\,{\AA} (see Ref.~\citenum{GJY+2014}
for details). Using this value, 
Eq.~\eqref{eq:Samp} predicts 
for pure ethanol: 
\begin{equation}
\mathcal{L}^{\rm th} = \frac{a^3}{C \eta A} \approx 10^3\, \text{L}/(\text{m}^2\cdot\text{h}\cdot\text{bar}),
\label{eq:Rth}
\end{equation}
where we used $\eta =1.1 \pm 0.1$\,mPa\,s for the viscosity of ethanol~\cite{GNV+2008} 
and $a$ is the effective radius of the tube given by $a \approx a_c - 2.5$\,{\AA} 
(taking into account the steric repulsion at the wall surface).
Note that the contribution of the Poiseuille-type dissipation inside the CNT
is negligible as compared to the entrance effect computed above, due to the large slip
at the CNT surface.

\subsubsection{\label{sec:osmotic}Osmotic pressure}

Beyond the linear dependence of the flux on the pressure, a 
more unexpected feature of the results in Fig.~\ref{fig:cnt_flux}(a)
is the existence of an offset in the pressure drop for
$x_w\ne 0$: for small
pressure drops, the ethanol flux is negative, i.e. directed towards
the ethanol-water mixture, and it becomes positive only above a threshold 
pressure drop.
This is the signature of an osmotic pressure expressed by the mixture across the membrane, suggesting that
the carbon membrane is semi-permeable to water.
This result is 
surprising because -- for most confinements -- the CNT membrane is 
in general permeable to both water and ethanol when they flow as pure
components.~\cite{Note1}
Accordingly the membranes become {``self semi-permeable''} to water 
due to a preferred adsorption of ethanol in nanoconfinement of the  carbon membrane, as compared
to the water.

Let us first explore more quantitatively the osmotic pressure.
For a membrane semi-permeable  to water, 
the ethanol flow
is expected to be proportional to $\Delta p - \Delta \Pi$ where $\Delta \Pi$
is the osmotic pressure due to the difference in water concentration
across the membrane. 
A simple thermodynamic formula for this osmotic pressure of 
the mixture yields~\cite{KR2008}
\begin{equation}
\Delta \Pi = - \dfrac{\rho_e N_{\rm A} k_{\rm B} T}{M_e} \sigma \ln\left(\gamma_e(1-x_w)\right),
\label{eq:pi}
\end{equation}
where $k_{\rm B}$ is the Boltzmann constant
and $\gamma_e$ is the activity coefficient of ethanol as a function of molar
fraction of water $x_w$.
The so-called reflection coefficient $\sigma$ accounts for the effect of
incomplete rejection of the solute (water in our case) through the
membrane,~\cite{Staverman1951,Fritz1986}
which at this stage is assumed to be unity.
Note that in the limit of a dilute solution, i.e. $N^B_w\to 0$ thus
$x_w\to 0$ and $\gamma_e\to 1$, this formula reduces to the van't Hoff law:
$\Delta \Pi=k_{\rm B}Tc_w$, where $c_w$ is the concentration of water
defined as the number of molecules per unit volume.

Figure~\ref{fig:cnt_flux}(b) shows the measured osmotic pressure
$\Delta\Pi$ in comparison with Eq.~\eqref{eq:pi}, as a function
of the initial value of $c_w$ in the bottom reservoir.
The osmotic pressure of the MD results is obtained from Fig.~\ref{fig:cnt_flux}(a),
by measuring the intersection point at $Q_e=0$.
The relation between concentration $c_w$ and molar fraction $x_w$ is 
$c_w=x_wN_{\rm A}/[M_e(1-x_w)/\rho_e+M_wx_w/\rho_w]$.
The values for the activity coefficient $\gamma_e$ 
used for the theoretical comparison of the osmotic pressure -- solid line in Fig.~\ref{fig:cnt_flux}(b) --
are shown in Fig.~\ref{fig:cnt_flux}(c). They are found  
to match very well the experimental values.~\cite{GP2007}
Altogether an excellent agreement is found between the theoretical prediction (with $\sigma=1$) 
and the MD results, assessing the semi-permeable character of the present CNT membrane to water
in the water-ethanol mixture. 
The rejection  
will be studied more quantitatively in the next section.

Another observation for the mixture is that the slope of
the $Q_e$ versus $\Delta p$ curve -- i.e. the hydrodynamic permeance to
ethanol -- appears to slightly decrease for  increasing water
concentration (with $\sim 10$\,\% variation). 
This suggests that the accumulated water molecules that appear near the
membrane in the steady state provide an additional resistance to ethanol
flow.~\cite{WNV+1985}

\subsubsection{Water selectivity}

\begin{figure}
\begin{center}
 \includegraphics[width=\linewidth]{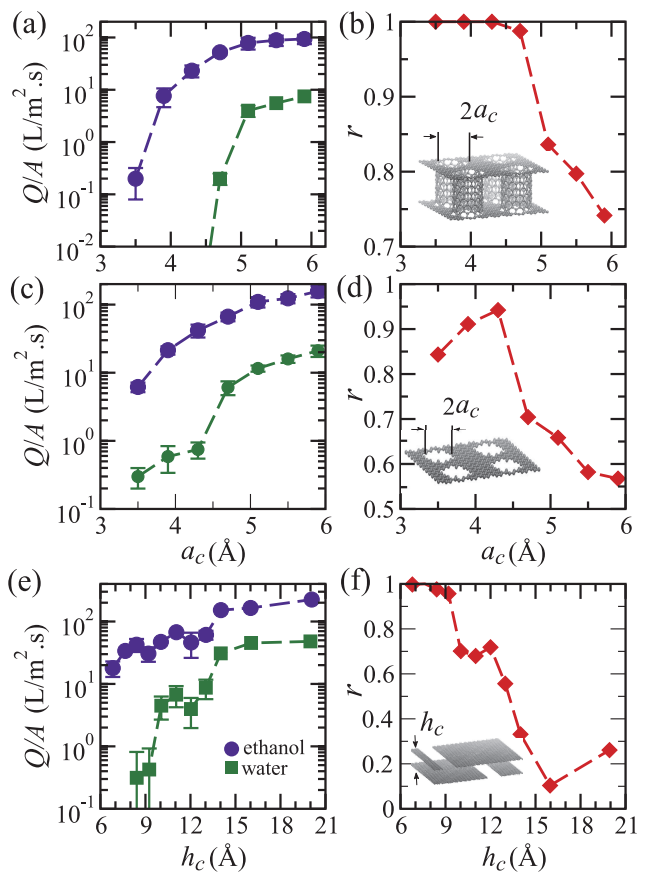}
\caption{
 Flow of ethanol-water mixture across carbon-based membranes
 of various types.
 Left: partial flux of ethanol and water per area $Q/A$ 
 for an ethanol-water mixture of $x_w=0.5$ flowing through
 (a) a CNT membrane and (c) a nanoporous graphene sheet, as a function of the
 pore size $a_c$,
under an applied pressure of $\Delta p=400$\,bar.
The corresponding flux through a multilayer graphene membrane 
with $\Delta p=800$\,bar
is plotted in panel (e), as a function of the  inter-layer distance
 $h_c$.
Right: Rejection coefficient $r$ (defined in Eq.~\eqref{eq:rejec}) for 
(b) the CNT membrane, (d) the nanoporous graphene,
and (f) the multilayer graphene membrane.
}
 \label{fig:main}
 \end{center}
\end{figure}

We now explore more exhaustively the water selectivity of the CNT membrane
as a function of the pore size.
To this end, we consider a water-ethanol mixture with an initial molar fraction
of $x_w=0.5$, i.e. a $50$\,\%-$50$\,\% mixture.
The flux of each component is measured across the membrane under
a given pressure-drop $\Delta p=400$\,bar,
and this procedure is repeated for various tube radii $a_c$ for the CNT membrane.
Results are reported in Fig. \ref{fig:main}(a).
As expected both fluxes (water and ethanol) increase 
for increasing tube radius.
Note that for the smaller tubes ($a_c = 3.5$, $3.9$ and $4.3$\,{\AA}),
no water molecule is recorded in the top reservoir during the total
duration of the simulation, corresponding to $40$\,ns.
In line with the observation in Sec.~\ref{sec:osmotic}, the flux of
ethanol is at least one order of magnitude larger than the flux of water.

In order to quantify the efficiency of the separation, we define the
rejection coefficient of the membrane as 
\begin{equation}
r = 1 - c \dfrac{Q_w}{Q_e},
\label{eq:rejec}
\end{equation}
where $c$ is defined as 
$c = M_e \rho_w (1 - x_w) / M_w \rho_e x_w$.
The prefactor $c$ is  such that the rejection coefficient $r$ is
equal to $1$ for a membrane completely impermeable to water and
is equal to $0$ for a membrane equally permeable to both ethanol and water.
As seen in Fig. \ref{fig:main}(b), $r$ is close to $1$ for 
$a_c \sim 4.7$\,{\AA}, and jumps down to $0.84$ for $a_c =5.1$\,{\AA}.
Note that $r$ is unity for pore radius below $4.7$\,{\AA} as the water
flux is negligible, predicting excellent separation performance for those radii.
The jump of the rejection coefficient between $4.7$ and $5.1$\,{\AA}
in radius echoes a previous result for water transport in CNT, in Ref.~\citenum{GYB+2016}, where
we showed that for in this radius range, 
 disjoining pressure effects reduce water adsorption in CNT.
This entropic effect may add up to the separation while having no effect
on ethanol permeability, in good
agreement with the present results.

\subsubsection{Affinity with the membrane}

\begin{figure}[t]
\begin{center}
\includegraphics[width=\linewidth]{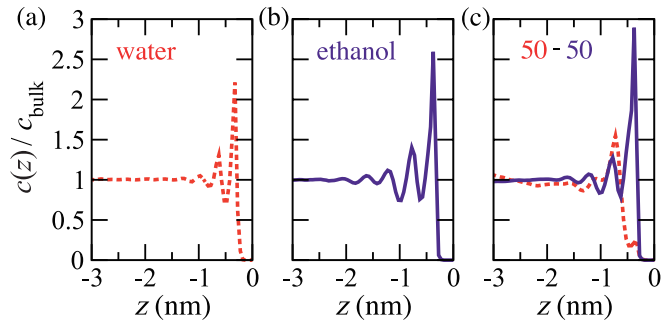}
\caption{
Concentration profiles of (a) pure water, (b) pure ethanol and (c) 
 mixture of $50\,\%$ water - $50\,\%$ ethanol, near a graphene sheet
 located at $z=0$.
 The profiles are normalized by their respective concentration
 in the bulk.
}
 \label{fig:adsorption}
\end{center}
\end{figure}

We now investigate the (molecular) mechanism underlying the
observed "self-semi-permeability." 
As we show here, the mechanism underlying the observed self semi-permeability
stems from the high affinity between the graphene
surface and ethanol molecules, in comparison to the  graphene-water interaction. 
This preferred affinity is highlighted by the detailed concentration profiles 
of water and ethanol near a graphene surface, as
shown in Fig.~\ref{fig:adsorption}.
Both pure ethanol and water liquids show a large absorption near the
graphene sheet, with the presence of a peak in the density profile
(Fig.~\ref{fig:adsorption}(a) and (b)). 
However in the case of a mixture, a higher affinity 
for the ethanol molecules is clearly observed in Fig.~\ref{fig:adsorption}(c),
with a strong peak of ethanol in the first layer near the
graphene sheet at $z=0$, while
most of the water molecules are displaced further away from the carbon
surface.
This preferred affinity of ethanol allows to rationalize the preferred
adsorption of ethanol in nanoconfining structure and the effective rejection
of water molecules, as we observe above in the membranes.
More into the details,
water is present in the second adsorption layer ($\geq0.5$\,nm away from the carbon surface).
This suggests
that the self-semi-permeability
requires confinement to be smaller than (roughly) two molecular layers. 
This is in agreement with the decrease of the
rejection coefficient in this range of confinement as observed in Fig.~\ref{fig:main}
for the three membranes considered here.

Furthermore, our findings are in agreement with experiments
reported in Ref.~\citenum{YYS+2013},
 showing the limited insertion of water into graphite oxide
in the presence of alcohol (methanol) in a mixture.
This points altogether to a robust physical mechanism,
and 
to the possibility of the separation using 
the GO membranes. Beyond the consequences on the osmotic behavior 
discussed here, this suggests
a rich behavior of the static  and structure properties of confined
mixtures, as pointed out in Ref.~\citenum{ZY2015}.


Finally, in order to
assess that the
effect is related to the preferred
adsorption of ethanol, we further checked the influence of the
solvent-carbon interaction strength. 
We performed additional simulations 
of water selectivity with three different force
fields (using OPLS-AA~\cite{JMT1996} instead of OPLS-UA for ethanol,
using SPC/E model ~\cite{BGS1987}
instead of TIP4P/2005 model for water, and using different Lennard-Jones
parameters for carbon atoms~\cite{SNC+2005}). 
These simulations gave qualitatively similar results as those shown here, 
with only  slight quantitative changes. This therefore supports the 
mechanism discussed in this section and
does confirm the robustness of the self-semi-permeability effect.

\subsection{Generalization to nanoporous  and multilayer graphene membranes}

\begin{figure}[t]
\begin{center}
 \includegraphics[width=0.7\linewidth]{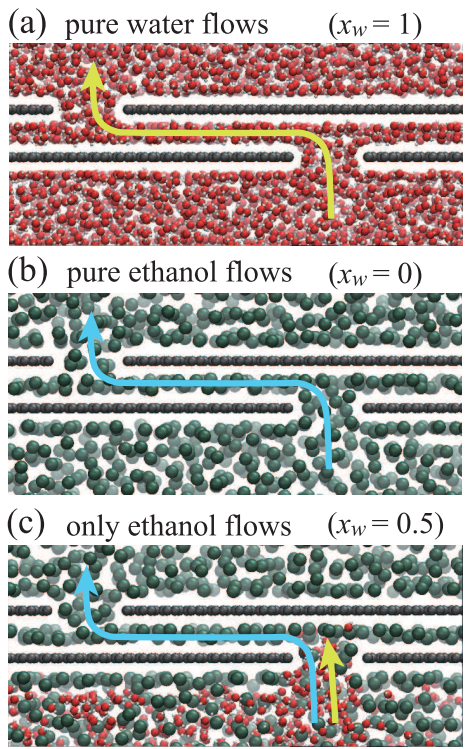}
\caption{
Snapshots of flows through the multilayer graphene membrane with
 $h_c=8.4$\,{\AA}, for (a) $x_w=1$ (with the top reservoir filled with water), (b) $x_w=0$, and (c) $x_w=0.5$.
The ethanol molecule is represented by a particle
at the position of CH$_2$. 
}
 \label{fig:flows}
 \end{center}
\end{figure}

Beyond the CNT membrane, 
the above  procedure was  applied to the various carbon membranes
under consideration: 
a graphene sheet pierced with circular
pores, reminiscent of the developing nanoporous graphene
membranes,~\cite{JCD2009} 
and 
multilayer graphene membranes, as depicted in Figs.~\ref{fig:geometry}(c) and (d).
The latter geometry is considered as a model of the porous
structure of the graphene-oxide (GO)
membranes.~\cite{NWJ+2012,XNZ+2015,HM2013,BXW+2016,ASM+2016,YB2016}
Overall the very same features are exhibited by {\it all considered carbon membranes}, namely: \\
(i) the membranes become {``self-semi-permeable''} to water in the presence
of water-ethanol mixtures, although both pure components pass freely
through them; this behavior is highlighted in Fig.~\ref{fig:flows} for
multilayer GO-like graphene membranes.\\
(ii) this semi-permeable character manifests itself in the expression of an osmotic
pressure, obeying the van't Hoff type expression, see Fig.~\ref{fig:cnt_flux}(b) for multilayer graphene membranes
and CNT membranes.\\
(iii) a size dependent water selectivity is measured, as highlighted in Fig.~\ref{fig:main}, confirming
semi-permeability for the smallest pore size. 

Overall an identical behavior for various confinement geometries points to a robust and generic 
mechanism. 
In line with the findings for the CNT membrane, the efficient separation of the
carbon membranes originates in the high affinity between carbon atoms and ethanol molecules, which leads to preferred carbon adsorption in the
nanoconfinement as compared to water.
Accordingly, the effect of separation  persists regardless of the details and geometry of the membrane, 
as long as the latter is made of carbon atom and
presents small pores (typically with a diameter $\le 1$\,nm).

The osmotic pressure $\Delta \Pi$ 
for the multilayer GO-like graphene
membranes is plotted in Fig.~\ref{fig:cnt_flux}(b).
For the case of $h_c=8.4$\,{\AA}, the osmotic pressure is again
in good agreement with the thermodynamic prediction given in
Eq.~\eqref{eq:pi}, which confirms  the semi-permeable character
of this membrane.
On the other hand, smaller values of $\Delta \Pi$ are obtained for the
GO-like membrane of $h_c=12$\,{\AA}. This implies that the 
rejection of water molecules is incomplete at this inter-layer distance.
Indeed, as shown by the dashed line in Fig.~\ref{fig:cnt_flux}(b),
the reduction of osmotic pressure is still captured by Eq.~\eqref{eq:pi}
with employing the value of the reflection coefficient smaller than
unity, i.e., $\sigma=0.7$.
 
As shown on Fig.~\ref{fig:main}, the three carbon membranes exhibit a similar rejection behavior,
with a rejection coefficient going from unity down for small pores to small rejection values as the typical size of
the pore (CNT diameter, pore size in graphene, or inter-layer gap for multilayer GO) bypasses a few
angtr\"oms. The range of rejection matches for all different membranes: typically the water exhibits high rejection
($r\approx 1$) for pore diameter below $10$\,{\AA}
for both CNTs and graphene pores (with a better performance for
CNTs) and for inter-layer gaps below $\sim 10$\,{\AA}
for multilayer GO. 
We note already at this stage that while it is difficult to fabricate macroscopic membranes of CNT and nanoporous graphene
with such specificities on the pore diameters, inter-layer gaps in this range are quite common for
macroscopic GO membranes.~\cite{YCW+2013} This is actually a very interesting feature for practical 
up-scaling of the process.

We also explored the influence of increasing the number of layers
in the multilayer GO membrane. As one may expect, we found that
selectivity increased with this number.
For the inter-layer distance $h_c=12$\,{\AA}, the
rejection coefficient reaches almost unity ($r>0.97$) with five-layer membrane,
with the flux of ethanol remaining in the same order of magnitude
as the two-layer membrane.
Furthermore, 
we note as a side remark that 
the measured values of flux for ethanol are
in the typical range of permeability estimate,~\cite{YB2016}
which  compares well with the experimental results for GO
membranes.~\cite{HM2013,XNZ+2015}

\section{Discussion}

Altogether, our results demonstrate that
carbon-based membranes can be used to separate very efficiently ethanol from water,
thereby suggesting their potential for membrane-based separation
of these two elements.~\cite{WDE2016}
In a very counter-intuitive way, these carbon-based membranes are shown to
be generally permeable to both liquids when considered as pure components, but
become semi-permeable to water for water-ethanol mixtures, as highlighted 
 in Fig.~\ref{fig:flows}.
This effective selectivity takes its origin 
in the high affinity of ethanol to these carbon membranes as compared
to water, an effect which is strongly enhanced for sufficiently small pore dimensions.
This separation mechanism is therefore robust, simple and quite independent 
of the geometry considered, as highlighted 
for the various types of carbon-based membranes considered in the
present study.
This effect leads to a rejection coefficient of water close to one
for carbon membranes with subnanometric pores, when in presence of water-ethanol
mixtures.
We further expect this separation process to apply not only to the ethanol-water
mixture, but also to any similar molecule, such as
methanol.~\cite{YYS+2013}
In addition, 
one may expect that the mechanisms behind the self-semi-permeability should
 persist to some extent in other hydrophobic (e.g. polymeric) nanoporous 
 membranes.~\cite{SYK+2014}

In order to highlight the potential of the membrane-based separation, 
let us quantify the energetics of the process. To put numbers, we consider a multilayer graphene 
membrane as a model system. Similar results are obtained with the two other types
of membranes, but this choice is particularly relevant because such graphene-oxide-like membranes 
are prone to easy scale-up.
To fix ideas we consider an multilayer carbon membrane with an inter-layer distance of $h_c = 8.4$\,{\AA}.
As shown in Fig.~\ref{fig:main}, this leads to nearly perfect water rejection, $r\simeq 1$.
A $50$\,\%-$50$\,\% water-ethanol mixture corresponds to an osmotic pressure $\Delta \Pi \approx 200$\,bar,
see Fig.~\ref{fig:cnt_flux}.
The ethanol flux under an applied pressure of $\Delta p=800$\,bar is found to be 
${Q/A} \sim 43$\,L/m$^2\cdot$s (see Fig.~\ref{fig:main}), while the water flux is negligible.
Assuming that the flux is proportional to 
$\Delta p - \Delta \Pi$, (see Sec.~\ref{sec:osmotic}), 
an ethanol flow rate of $Q_{e}\sim 3.5$\,L per second for a $1$\,m$^2$ membrane will be driven
under an applied pressure of $\Delta p = 250$\,bar. 
The corresponding required power is accordingly 
${\cal P} = Q_{e}\times \Delta p \sim 88\,$ kW.  
The cost for separating, say, $1$\,L of water-ethanol, is then $\sim 25$\,kJ.
This energy cost is to be compared to the thermodynamic limit for the energy cost of separating such
a mixture, which is $17$\,kJ (see e.g. Ref.~\citenum{AH1997}).
Furthermore the energy required for boiling in a typical 
azeotropic distillation process, based on an extrapolation of 
reported value,~\cite{PS1983} is approximately $3$\,MJ, which is two orders of magnitude
larger than that estimated for the carbon multilayer membrane.
Similar numbers are expected for the other CNT and nanoporous graphene
membranes.

These predictions are accordingly highly attractive as an alternative solution
for water-ethanol separation. 
More specifically the results for the CNT and nanoporous membranes may
certainly suggest high expectations, and the fabrication of carbon nanotube
membranes,~\cite{HPW+2006, MCA+2005} carbon 
nanotube-mixed matrix membranes,~\cite{IGS+2009} ultra thin nanoporous graphene
membranes,~\cite{BBN+2008} have been reported in the recent literature. 
But the practical scaling-up of such
membranes up to square meters still remains a technological challenge. 
In contrast, graphene-oxides membranes,
which are intrinsically large scale and easy to fabricate,~\cite{DSZ+2007,LSZ+2013,KYY+2013}
makes such layered carbon membranes a far more plausible candidate
to highlight the present effect. 
We believe that the potentially huge reduction in energy cost offered by
the present membrane-based process makes it a serious candidate for water-ethanol
separation at large scales.

\begin{acknowledgments}
This research was supported by the European Research Council
program {\it Micromegas} project.
It was also granted access to the HPC resources of MesoPSL financed
by the Region Ile de France and the project Equip@Meso (reference
ANR-10-EQPX-29-01) of the programme Investissements d'Avenir supervised
by the Agence Nationale de la Recherche (ANR).
\end{acknowledgments}


\begin{thebibliography}{57}%
\makeatletter
\providecommand \@ifxundefined [1]{%
 \@ifx{#1\undefined}
}%
\providecommand \@ifnum [1]{%
 \ifnum #1\expandafter \@firstoftwo
 \else \expandafter \@secondoftwo
 \fi
}%
\providecommand \@ifx [1]{%
 \ifx #1\expandafter \@firstoftwo
 \else \expandafter \@secondoftwo
 \fi
}%
\providecommand \natexlab [1]{#1}%
\providecommand \enquote  [1]{``#1''}%
\providecommand \bibnamefont  [1]{#1}%
\providecommand \bibfnamefont [1]{#1}%
\providecommand \citenamefont [1]{#1}%
\providecommand \href@noop [0]{\@secondoftwo}%
\providecommand \href [0]{\begingroup \@sanitize@url \@href}%
\providecommand \@href[1]{\@@startlink{#1}\@@href}%
\providecommand \@@href[1]{\endgroup#1\@@endlink}%
\providecommand \@sanitize@url [0]{\catcode `\\12\catcode `\$12\catcode
  `\&12\catcode `\#12\catcode `\^12\catcode `\_12\catcode `\%12\relax}%
\providecommand \@@startlink[1]{}%
\providecommand \@@endlink[0]{}%
\providecommand \url  [0]{\begingroup\@sanitize@url \@url }%
\providecommand \@url [1]{\endgroup\@href {#1}{\urlprefix }}%
\providecommand \urlprefix  [0]{URL }%
\providecommand \Eprint [0]{\href }%
\providecommand \doibase [0]{http://dx.doi.org/}%
\providecommand \selectlanguage [0]{\@gobble}%
\providecommand \bibinfo  [0]{\@secondoftwo}%
\providecommand \bibfield  [0]{\@secondoftwo}%
\providecommand \translation [1]{[#1]}%
\providecommand \BibitemOpen [0]{}%
\providecommand \bibitemStop [0]{}%
\providecommand \bibitemNoStop [0]{.\EOS\space}%
\providecommand \EOS [0]{\spacefactor3000\relax}%
\providecommand \BibitemShut  [1]{\csname bibitem#1\endcsname}%
\let\auto@bib@innerbib\@empty
\bibitem [{\citenamefont {Balat}\ and\ \citenamefont {Balat}(2009)}]{BB2009}%
  \BibitemOpen
  \bibfield  {author} {\bibinfo {author} {\bibfnamefont {M.}~\bibnamefont
  {Balat}}\ and\ \bibinfo {author} {\bibfnamefont {H.}~\bibnamefont {Balat}},\
  }\bibfield  {title} {\enquote {\bibinfo {title} {Recent trends in global
  production and utilization of bio-ethanol fuel},}\ }\href@noop {} {\bibfield
  {journal} {\bibinfo  {journal} {Appl. Energ.}\ }\textbf {\bibinfo {volume}
  {86}},\ \bibinfo {pages} {2273--2282} (\bibinfo {year} {2009})}\BibitemShut
  {NoStop}%
\bibitem [{\citenamefont {Hahn-H{\"a}gerdal}\ \emph {et~al.}(2006)\citenamefont
  {Hahn-H{\"a}gerdal}, \citenamefont {Galbe}, \citenamefont {Gorwa-Grauslund},
  \citenamefont {Lid{\'e}n},\ and\ \citenamefont {Zacchi}}]{HGG+2006}%
  \BibitemOpen
  \bibfield  {author} {\bibinfo {author} {\bibfnamefont {B.}~\bibnamefont
  {Hahn-H{\"a}gerdal}}, \bibinfo {author} {\bibfnamefont {M.}~\bibnamefont
  {Galbe}}, \bibinfo {author} {\bibfnamefont {M.~F.}\ \bibnamefont
  {Gorwa-Grauslund}}, \bibinfo {author} {\bibfnamefont {G.}~\bibnamefont
  {Lid{\'e}n}}, \ and\ \bibinfo {author} {\bibfnamefont {G.}~\bibnamefont
  {Zacchi}},\ }\bibfield  {title} {\enquote {\bibinfo {title} {Bio-ethanol--the
  fuel of tomorrow from the residues of today},}\ }\href@noop {} {\bibfield
  {journal} {\bibinfo  {journal} {Trends. Biotechnol.}\ }\textbf {\bibinfo
  {volume} {24}},\ \bibinfo {pages} {549--556} (\bibinfo {year}
  {2006})}\BibitemShut {NoStop}%
\bibitem [{\citenamefont {Huang}\ \emph {et~al.}(2008)\citenamefont {Huang},
  \citenamefont {Ramaswamy}, \citenamefont {Tschirner},\ and\ \citenamefont
  {Ramarao}}]{HRT+2008}%
  \BibitemOpen
  \bibfield  {author} {\bibinfo {author} {\bibfnamefont {H.-J.}\ \bibnamefont
  {Huang}}, \bibinfo {author} {\bibfnamefont {S.}~\bibnamefont {Ramaswamy}},
  \bibinfo {author} {\bibfnamefont {U.~W.}\ \bibnamefont {Tschirner}}, \ and\
  \bibinfo {author} {\bibfnamefont {B.~V.}\ \bibnamefont {Ramarao}},\
  }\bibfield  {title} {\enquote {\bibinfo {title} {A review of separation
  technologies in current and future biorefineries},}\ }\href@noop {}
  {\bibfield  {journal} {\bibinfo  {journal} {Sep. Purif. Technol.}\ }\textbf
  {\bibinfo {volume} {62}},\ \bibinfo {pages} {1--21} (\bibinfo {year}
  {2008})}\BibitemShut {NoStop}%
\bibitem [{\citenamefont {Mulder}\ \emph {et~al.}(1983)\citenamefont {Mulder},
  \citenamefont {Hendrickman}, \citenamefont {Hegeman},\ and\ \citenamefont
  {Smolders}}]{MHH+1983}%
  \BibitemOpen
  \bibfield  {author} {\bibinfo {author} {\bibfnamefont {M.~H.~V.}\
  \bibnamefont {Mulder}}, \bibinfo {author} {\bibfnamefont {J.~O.}\
  \bibnamefont {Hendrickman}}, \bibinfo {author} {\bibfnamefont
  {H.}~\bibnamefont {Hegeman}}, \ and\ \bibinfo {author} {\bibfnamefont
  {C.~A.}\ \bibnamefont {Smolders}},\ }\bibfield  {title} {\enquote {\bibinfo
  {title} {Ethanol--water separation by pervaporation},}\ }\href@noop {}
  {\bibfield  {journal} {\bibinfo  {journal} {J. Membrane Sci.}\ }\textbf
  {\bibinfo {volume} {16}},\ \bibinfo {pages} {269--284} (\bibinfo {year}
  {1983})}\BibitemShut {NoStop}%
\bibitem [{\citenamefont {Sano}\ \emph {et~al.}(1994)\citenamefont {Sano},
  \citenamefont {Yanagishita}, \citenamefont {Kiyozumi}, \citenamefont
  {Mizukami},\ and\ \citenamefont {Haraya}}]{SYK+1994}%
  \BibitemOpen
  \bibfield  {author} {\bibinfo {author} {\bibfnamefont {T.}~\bibnamefont
  {Sano}}, \bibinfo {author} {\bibfnamefont {H.}~\bibnamefont {Yanagishita}},
  \bibinfo {author} {\bibfnamefont {Y.}~\bibnamefont {Kiyozumi}}, \bibinfo
  {author} {\bibfnamefont {F.}~\bibnamefont {Mizukami}}, \ and\ \bibinfo
  {author} {\bibfnamefont {K.}~\bibnamefont {Haraya}},\ }\bibfield  {title}
  {\enquote {\bibinfo {title} {Separation of ethanol/water mixture by
  silicalite membrane on pervaporation},}\ }\href@noop {} {\bibfield  {journal}
  {\bibinfo  {journal} {J. Membrane Sci.}\ }\textbf {\bibinfo {volume} {95}},\
  \bibinfo {pages} {221--228} (\bibinfo {year} {1994})}\BibitemShut {NoStop}%
\bibitem [{\citenamefont {Mulder}, \citenamefont {Franken},\ and\ \citenamefont
  {Smolders}(1985)}]{MFS1985}%
  \BibitemOpen
  \bibfield  {author} {\bibinfo {author} {\bibfnamefont {M.~H.~V.}\
  \bibnamefont {Mulder}}, \bibinfo {author} {\bibfnamefont {A.~C.~M.}\
  \bibnamefont {Franken}}, \ and\ \bibinfo {author} {\bibfnamefont {C.~A.}\
  \bibnamefont {Smolders}},\ }\bibfield  {title} {\enquote {\bibinfo {title}
  {On the mechanism of separation of ethanol/water mixtures by pervaporation
  {II}. {E}xperimental concentration profiles},}\ }\href@noop {} {\bibfield
  {journal} {\bibinfo  {journal} {J. Membrane Sci.}\ }\textbf {\bibinfo
  {volume} {23}},\ \bibinfo {pages} {41--58} (\bibinfo {year}
  {1985})}\BibitemShut {NoStop}%
\bibitem [{\citenamefont {Nomura}, \citenamefont {Yamaguchi},\ and\
  \citenamefont {Nakao}(1998)}]{NYN1998}%
  \BibitemOpen
  \bibfield  {author} {\bibinfo {author} {\bibfnamefont {M.}~\bibnamefont
  {Nomura}}, \bibinfo {author} {\bibfnamefont {T.}~\bibnamefont {Yamaguchi}}, \
  and\ \bibinfo {author} {\bibfnamefont {S.-I.}\ \bibnamefont {Nakao}},\
  }\bibfield  {title} {\enquote {\bibinfo {title} {Ethanol/water transport
  through silicalite membranes},}\ }\href@noop {} {\bibfield  {journal}
  {\bibinfo  {journal} {J. Membrane Sci.}\ }\textbf {\bibinfo {volume} {144}},\
  \bibinfo {pages} {161--171} (\bibinfo {year} {1998})}\BibitemShut {NoStop}%
\bibitem [{\citenamefont {Semiat}(2008)}]{Semiat2008}%
  \BibitemOpen
  \bibfield  {author} {\bibinfo {author} {\bibfnamefont {R.}~\bibnamefont
  {Semiat}},\ }\bibfield  {title} {\enquote {\bibinfo {title} {Energy issues in
  desalination processes},}\ }\href@noop {} {\bibfield  {journal} {\bibinfo
  {journal} {Environ. Sci. Technol.}\ }\textbf {\bibinfo {volume} {42}},\
  \bibinfo {pages} {8193--8201} (\bibinfo {year} {2008})}\BibitemShut {NoStop}%
\bibitem [{\citenamefont {Sablani}\ \emph {et~al.}(2001)\citenamefont
  {Sablani}, \citenamefont {Goosen}, \citenamefont {Al-Belushi},\ and\
  \citenamefont {Wilf}}]{SGA+2001}%
  \BibitemOpen
  \bibfield  {author} {\bibinfo {author} {\bibfnamefont {S.~S.}\ \bibnamefont
  {Sablani}}, \bibinfo {author} {\bibfnamefont {M.~F.~A.}\ \bibnamefont
  {Goosen}}, \bibinfo {author} {\bibfnamefont {R.}~\bibnamefont {Al-Belushi}},
  \ and\ \bibinfo {author} {\bibfnamefont {M.}~\bibnamefont {Wilf}},\
  }\bibfield  {title} {\enquote {\bibinfo {title} {Concentration polarization
  in ultrafiltration and reverse osmosis: a critical review},}\ }\href@noop {}
  {\bibfield  {journal} {\bibinfo  {journal} {Desalination}\ }\textbf {\bibinfo
  {volume} {141}},\ \bibinfo {pages} {269--289} (\bibinfo {year}
  {2001})}\BibitemShut {NoStop}%
\bibitem [{\citenamefont {Ismail}\ \emph {et~al.}(2009)\citenamefont {Ismail},
  \citenamefont {Goh}, \citenamefont {Sanip},\ and\ \citenamefont
  {Aziz}}]{IGS+2009}%
  \BibitemOpen
  \bibfield  {author} {\bibinfo {author} {\bibfnamefont {A.~F.}\ \bibnamefont
  {Ismail}}, \bibinfo {author} {\bibfnamefont {P.~S.}\ \bibnamefont {Goh}},
  \bibinfo {author} {\bibfnamefont {S.~M.}\ \bibnamefont {Sanip}}, \ and\
  \bibinfo {author} {\bibfnamefont {M.}~\bibnamefont {Aziz}},\ }\bibfield
  {title} {\enquote {\bibinfo {title} {Transport and separation properties of
  carbon nanotube-mixed matrix membrane},}\ }\href@noop {} {\bibfield
  {journal} {\bibinfo  {journal} {Sep. Purif. Technol.}\ }\textbf {\bibinfo
  {volume} {70}},\ \bibinfo {pages} {12--26} (\bibinfo {year}
  {2009})}\BibitemShut {NoStop}%
\bibitem [{\citenamefont {Lee}, \citenamefont {Arnot},\ and\ \citenamefont
  {Mattia}(2011)}]{LAM2011}%
  \BibitemOpen
  \bibfield  {author} {\bibinfo {author} {\bibfnamefont {K.~P.}\ \bibnamefont
  {Lee}}, \bibinfo {author} {\bibfnamefont {T.~C.}\ \bibnamefont {Arnot}}, \
  and\ \bibinfo {author} {\bibfnamefont {D.}~\bibnamefont {Mattia}},\
  }\bibfield  {title} {\enquote {\bibinfo {title} {A review of reverse osmosis
  membrane materials for desalination--development to date and future
  potential},}\ }\href@noop {} {\bibfield  {journal} {\bibinfo  {journal} {J.
  Membrane Sci.}\ }\textbf {\bibinfo {volume} {370}},\ \bibinfo {pages} {1--22}
  (\bibinfo {year} {2011})}\BibitemShut {NoStop}%
\bibitem [{\citenamefont {Fritzmann}\ \emph {et~al.}(2007)\citenamefont
  {Fritzmann}, \citenamefont {L{\"o}wenberg}, \citenamefont {Wintgens},\ and\
  \citenamefont {Melin}}]{FLW+2007}%
  \BibitemOpen
  \bibfield  {author} {\bibinfo {author} {\bibfnamefont {C.}~\bibnamefont
  {Fritzmann}}, \bibinfo {author} {\bibfnamefont {J.}~\bibnamefont
  {L{\"o}wenberg}}, \bibinfo {author} {\bibfnamefont {T.}~\bibnamefont
  {Wintgens}}, \ and\ \bibinfo {author} {\bibfnamefont {T.}~\bibnamefont
  {Melin}},\ }\bibfield  {title} {\enquote {\bibinfo {title} {State-of-the-art
  of reverse osmosis desalination},}\ }\href@noop {} {\bibfield  {journal}
  {\bibinfo  {journal} {Desalination}\ }\textbf {\bibinfo {volume} {216}},\
  \bibinfo {pages} {1--76} (\bibinfo {year} {2007})}\BibitemShut {NoStop}%
\bibitem{Note1}
 {Note that we find that all carbon membranes are permeable to pure water, except
 the CNT membrane for a specific range of pore radii $a_c\in[4.3-4.7]$~{\AA}. See Ref.~\citenum{GYB+2016}}.
\bibitem [{\citenamefont {Plimpton}(1995)}]{Plimpton1995}%
  \BibitemOpen
  \bibfield  {author} {\bibinfo {author} {\bibfnamefont {S.}~\bibnamefont
  {Plimpton}},\ }\bibfield  {title} {\enquote {\bibinfo {title} {Fast parallel
  algorithms for short-range molecular dynamics},}\ }\href@noop {} {\bibfield
  {journal} {\bibinfo  {journal} {J. Comput. Phys.}\ }\textbf {\bibinfo
  {volume} {117}},\ \bibinfo {pages} {1--19} (\bibinfo {year}
  {1995})}\BibitemShut {NoStop}%
\bibitem [{LAM()}]{LAMMPS}%
  \BibitemOpen
  \href@noop {} {}\bibinfo {howpublished} {See \url{http://lammps.sandia.gov}
  for the code}\BibitemShut {NoStop}%
\bibitem [{\citenamefont {Abascal}\ and\ \citenamefont {Vega}(2005)}]{AV2005}%
  \BibitemOpen
  \bibfield  {author} {\bibinfo {author} {\bibfnamefont {J.~L.~F.}\
  \bibnamefont {Abascal}}\ and\ \bibinfo {author} {\bibfnamefont
  {C.}~\bibnamefont {Vega}},\ }\bibfield  {title} {\enquote {\bibinfo {title}
  {A general purpose model for the condensed phases of water: {TIP4P/2005}},}\
  }\href@noop {} {\bibfield  {journal} {\bibinfo  {journal} {J. Chem. Phys.}\
  }\textbf {\bibinfo {volume} {123}},\ \bibinfo {pages} {234505} (\bibinfo
  {year} {2005})}\BibitemShut {NoStop}%
\bibitem [{\citenamefont {Jorgensen}, \citenamefont {Madura},\ and\
  \citenamefont {Swenson}(1984)}]{JMS1984}%
  \BibitemOpen
  \bibfield  {author} {\bibinfo {author} {\bibfnamefont {W.~L.}\ \bibnamefont
  {Jorgensen}}, \bibinfo {author} {\bibfnamefont {J.~D.}\ \bibnamefont
  {Madura}}, \ and\ \bibinfo {author} {\bibfnamefont {C.~J.}\ \bibnamefont
  {Swenson}},\ }\bibfield  {title} {\enquote {\bibinfo {title} {Optimized
  intermolecular potential functions for liquid hydrocarbons},}\ }\href@noop {}
  {\bibfield  {journal} {\bibinfo  {journal} {J. A}\ }\textbf {\bibinfo
  {volume} {106}},\ \bibinfo {pages} {6638--6646} (\bibinfo {year}
  {1984})}\BibitemShut {NoStop}%
\bibitem [{\citenamefont {Jorgensen}(1986)}]{Jorgensen1986}%
  \BibitemOpen
  \bibfield  {author} {\bibinfo {author} {\bibfnamefont {W.~L.}\ \bibnamefont
  {Jorgensen}},\ }\bibfield  {title} {\enquote {\bibinfo {title} {Optimized
  intermolecular potential functions for liquid alcohols},}\ }\href@noop {}
  {\bibfield  {journal} {\bibinfo  {journal} {J. Chem. Phys.}\ }\textbf
  {\bibinfo {volume} {90}},\ \bibinfo {pages} {1276--1284} (\bibinfo {year}
  {1986})}\BibitemShut {NoStop}%
\bibitem [{\citenamefont {Cornell}\ \emph {et~al.}(1995)\citenamefont
  {Cornell}, \citenamefont {Cieplak}, \citenamefont {Bayly}, \citenamefont
  {Gould}, \citenamefont {Merz}, \citenamefont {Ferguson}, \citenamefont
  {Spellmeyer}, \citenamefont {Fox}, \citenamefont {Caldwell},\ and\
  \citenamefont {Kollman}}]{CCB+1995}%
  \BibitemOpen
  \bibfield  {author} {\bibinfo {author} {\bibfnamefont {W.~D.}\ \bibnamefont
  {Cornell}}, \bibinfo {author} {\bibfnamefont {P.}~\bibnamefont {Cieplak}},
  \bibinfo {author} {\bibfnamefont {C.~I.}\ \bibnamefont {Bayly}}, \bibinfo
  {author} {\bibfnamefont {I.~R.}\ \bibnamefont {Gould}}, \bibinfo {author}
  {\bibfnamefont {K.~M.}\ \bibnamefont {Merz}}, \bibinfo {author}
  {\bibfnamefont {D.~M.}\ \bibnamefont {Ferguson}}, \bibinfo {author}
  {\bibfnamefont {D.~C.}\ \bibnamefont {Spellmeyer}}, \bibinfo {author}
  {\bibfnamefont {T.}~\bibnamefont {Fox}}, \bibinfo {author} {\bibfnamefont
  {J.~W.}\ \bibnamefont {Caldwell}}, \ and\ \bibinfo {author} {\bibfnamefont
  {P.~A.}\ \bibnamefont {Kollman}},\ }\bibfield  {title} {\enquote {\bibinfo
  {title} {A second generation force field for the simulation of proteins,
  nucleic acids, and organic molecules},}\ }\href@noop {} {\bibfield  {journal}
  {\bibinfo  {journal} {J. Am. Chem. Soc.}\ }\textbf {\bibinfo {volume}
  {117}},\ \bibinfo {pages} {5179--5197} (\bibinfo {year} {1995})}\BibitemShut
  {NoStop}%
\bibitem [{\citenamefont {Alexiadis}\ and\ \citenamefont
  {Kassinos}(2008)}]{AK2008}%
  \BibitemOpen
  \bibfield  {author} {\bibinfo {author} {\bibfnamefont {A.}~\bibnamefont
  {Alexiadis}}\ and\ \bibinfo {author} {\bibfnamefont {S.}~\bibnamefont
  {Kassinos}},\ }\bibfield  {title} {\enquote {\bibinfo {title} {Molecular
  simulation of water in carbon nanotubes},}\ }\href@noop {} {\bibfield
  {journal} {\bibinfo  {journal} {Chem. Rev.}\ }\textbf {\bibinfo {volume}
  {108}},\ \bibinfo {pages} {5014--5034} (\bibinfo {year} {2008})}\BibitemShut
  {NoStop}%
\bibitem [{\citenamefont {Thomas}\ and\ \citenamefont
  {McGaughey}(2009)}]{TM2009}%
  \BibitemOpen
  \bibfield  {author} {\bibinfo {author} {\bibfnamefont {J.~A.}\ \bibnamefont
  {Thomas}}\ and\ \bibinfo {author} {\bibfnamefont {A.~J.~H.}\ \bibnamefont
  {McGaughey}},\ }\bibfield  {title} {\enquote {\bibinfo {title} {Water flow in
  carbon nanotubes: transition to subcontinuum transport},}\ }\href@noop {}
  {\bibfield  {journal} {\bibinfo  {journal} {Phys. Rev. Lett.}\ }\textbf
  {\bibinfo {volume} {102}},\ \bibinfo {pages} {184502} (\bibinfo {year}
  {2009})}\BibitemShut {NoStop}%
\bibitem [{\citenamefont {Werder}\ \emph {et~al.}(2003)\citenamefont {Werder},
  \citenamefont {Walther}, \citenamefont {Jaffe}, \citenamefont {Halicioglu},\
  and\ \citenamefont {Koumoutsakos}}]{WWJ+2003}%
  \BibitemOpen
  \bibfield  {author} {\bibinfo {author} {\bibfnamefont {T.}~\bibnamefont
  {Werder}}, \bibinfo {author} {\bibfnamefont {J.~H.}\ \bibnamefont {Walther}},
  \bibinfo {author} {\bibfnamefont {R.~L.}\ \bibnamefont {Jaffe}}, \bibinfo
  {author} {\bibfnamefont {T.}~\bibnamefont {Halicioglu}}, \ and\ \bibinfo
  {author} {\bibfnamefont {P.}~\bibnamefont {Koumoutsakos}},\ }\bibfield
  {title} {\enquote {\bibinfo {title} {On the water-carbon interaction for use
  in molecular dynamics simulations of graphite and carbon nanotubes},}\
  }\href@noop {} {\bibfield  {journal} {\bibinfo  {journal} {J. Phys. Chem. B}\
  }\textbf {\bibinfo {volume} {107}},\ \bibinfo {pages} {1345--1352} (\bibinfo
  {year} {2003})}\BibitemShut {NoStop}%
\bibitem [{\citenamefont {Berendsen}\ \emph {et~al.}(1984)\citenamefont
  {Berendsen}, \citenamefont {Postma}, \citenamefont {van Gunsteren},
  \citenamefont {DiNola},\ and\ \citenamefont {Haak}}]{BPG+1984}%
  \BibitemOpen
  \bibfield  {author} {\bibinfo {author} {\bibfnamefont {H.~J.~C.}\
  \bibnamefont {Berendsen}}, \bibinfo {author} {\bibfnamefont {J.~P.~M.}\
  \bibnamefont {Postma}}, \bibinfo {author} {\bibfnamefont {W.~F.}\
  \bibnamefont {van Gunsteren}}, \bibinfo {author} {\bibfnamefont
  {A.}~\bibnamefont {DiNola}}, \ and\ \bibinfo {author} {\bibfnamefont {J.~R.}\
  \bibnamefont {Haak}},\ }\bibfield  {title} {\enquote {\bibinfo {title}
  {Molecular dynamics with coupling to an external bath},}\ }\href@noop {}
  {\bibfield  {journal} {\bibinfo  {journal} {J. Chem. Phys.}\ }\textbf
  {\bibinfo {volume} {81}},\ \bibinfo {pages} {3684--3690} (\bibinfo {year}
  {1984})}\BibitemShut {NoStop}%
\bibitem [{\citenamefont {Thomas}\ and\ \citenamefont {Corry}(2015)}]{TC2015}%
  \BibitemOpen
  \bibfield  {author} {\bibinfo {author} {\bibfnamefont {M.}~\bibnamefont
  {Thomas}}\ and\ \bibinfo {author} {\bibfnamefont {B.}~\bibnamefont {Corry}},\
  }\bibfield  {title} {\enquote {\bibinfo {title} {Thermostat choice
  significantly influences water flow rates in molecular dynamics studies of
  carbon nanotubes},}\ }\href@noop {} {\bibfield  {journal} {\bibinfo
  {journal} {Microfluid. Nanofluid.}\ }\textbf {\bibinfo {volume} {18}},\
  \bibinfo {pages} {41--47} (\bibinfo {year} {2015})}\BibitemShut {NoStop}%
\bibitem [{\citenamefont {Green}\ and\ \citenamefont {Perry}(2007)}]{GP2007}%
  \BibitemOpen
  \bibfield  {author} {\bibinfo {author} {\bibfnamefont {D.~W.}\ \bibnamefont
  {Green}}\ and\ \bibinfo {author} {\bibfnamefont {R.~H.}\ \bibnamefont
  {Perry}},\ }\href@noop {} {\emph {\bibinfo {title} {Perry's chemical
  engineers' handbook}}},\ \bibinfo {edition} {8th}\ ed.\ (\bibinfo
  {publisher} {McGraw-Hill},\ \bibinfo {year} {2007})\BibitemShut {NoStop}%
\bibitem [{\citenamefont {Yoshida}\ and\ \citenamefont
  {Bocquet}(2016)}]{YB2016}%
  \BibitemOpen
  \bibfield  {author} {\bibinfo {author} {\bibfnamefont {H.}~\bibnamefont
  {Yoshida}}\ and\ \bibinfo {author} {\bibfnamefont {L.}~\bibnamefont
  {Bocquet}},\ }\bibfield  {title} {\enquote {\bibinfo {title} {Labyrinthine
  water flow across multilayer graphene-based membranes: molecular dynamics
  versus continuum predictions},}\ }\href@noop {} {\bibfield  {journal}
  {\bibinfo  {journal} {J. Chem. Phys.}\ }\textbf {\bibinfo {volume} {144}},\
  \bibinfo {pages} {234701} (\bibinfo {year} {2016})}\BibitemShut {NoStop}%
\bibitem [{\citenamefont {Falk}\ \emph {et~al.}(2012)\citenamefont {Falk},
  \citenamefont {Sedlmeier}, \citenamefont {Joly}, \citenamefont {Netz},\ and\
  \citenamefont {Bocquet}}]{FSJ+2012}%
  \BibitemOpen
  \bibfield  {author} {\bibinfo {author} {\bibfnamefont {K.}~\bibnamefont
  {Falk}}, \bibinfo {author} {\bibfnamefont {F.}~\bibnamefont {Sedlmeier}},
  \bibinfo {author} {\bibfnamefont {L.}~\bibnamefont {Joly}}, \bibinfo {author}
  {\bibfnamefont {R.~R.}\ \bibnamefont {Netz}}, \ and\ \bibinfo {author}
  {\bibfnamefont {L.}~\bibnamefont {Bocquet}},\ }\bibfield  {title} {\enquote
  {\bibinfo {title} {Ultralow liquid/solid friction in carbon nanotubes:
  comprehensive theory for alcohols, alkanes, {OMCTS}, and water},}\
  }\href@noop {} {\bibfield  {journal} {\bibinfo  {journal} {Langmuir}\
  }\textbf {\bibinfo {volume} {28}},\ \bibinfo {pages} {14261--14272} (\bibinfo
  {year} {2012})}\BibitemShut {NoStop}%
\bibitem [{\citenamefont {Sisan}\ and\ \citenamefont {Lichter}(2011)}]{SL2011}%
  \BibitemOpen
  \bibfield  {author} {\bibinfo {author} {\bibfnamefont {T.~B.}\ \bibnamefont
  {Sisan}}\ and\ \bibinfo {author} {\bibfnamefont {S.}~\bibnamefont
  {Lichter}},\ }\bibfield  {title} {\enquote {\bibinfo {title} {The end of
  nanochannels},}\ }\href@noop {} {\bibfield  {journal} {\bibinfo  {journal}
  {Microfluid. Nanofluid.}\ }\textbf {\bibinfo {volume} {11}},\ \bibinfo
  {pages} {787--791} (\bibinfo {year} {2011})}\BibitemShut {NoStop}%
\bibitem [{\citenamefont {Sampson}(1891)}]{Sampson1891}%
  \BibitemOpen
  \bibfield  {author} {\bibinfo {author} {\bibfnamefont {R.~A.}\ \bibnamefont
  {Sampson}},\ }\bibfield  {title} {\enquote {\bibinfo {title} {On {S}tokes's
  current function},}\ }\href@noop {} {\bibfield  {journal} {\bibinfo
  {journal} {Phil. Trans. R. Soc. A}\ }\textbf {\bibinfo {volume} {182}},\
  \bibinfo {pages} {449--518} (\bibinfo {year} {1891})}\BibitemShut {NoStop}%
\bibitem [{\citenamefont {Gravelle}\ \emph {et~al.}(2014)\citenamefont
  {Gravelle}, \citenamefont {Joly}, \citenamefont {Ybert},\ and\ \citenamefont
  {Bocquet}}]{GJY+2014}%
  \BibitemOpen
  \bibfield  {author} {\bibinfo {author} {\bibfnamefont {S.}~\bibnamefont
  {Gravelle}}, \bibinfo {author} {\bibfnamefont {L.}~\bibnamefont {Joly}},
  \bibinfo {author} {\bibfnamefont {C.}~\bibnamefont {Ybert}}, \ and\ \bibinfo
  {author} {\bibfnamefont {L.}~\bibnamefont {Bocquet}},\ }\bibfield  {title}
  {\enquote {\bibinfo {title} {Large permeabilities of hourglass nanopores:
  from hydrodynamics to single file transport},}\ }\href@noop {} {\bibfield
  {journal} {\bibinfo  {journal} {J. Chem. Phys.}\ }\textbf {\bibinfo {volume}
  {141}},\ \bibinfo {pages} {18C526} (\bibinfo {year} {2014})}\BibitemShut
  {NoStop}%
\bibitem [{\citenamefont {Guevara-Carrion}\ \emph {et~al.}(2008)\citenamefont
  {Guevara-Carrion}, \citenamefont {Nieto-Draghi}, \citenamefont {Vrabec},\
  and\ \citenamefont {Hasse}}]{GNV+2008}%
  \BibitemOpen
  \bibfield  {author} {\bibinfo {author} {\bibfnamefont {G.}~\bibnamefont
  {Guevara-Carrion}}, \bibinfo {author} {\bibfnamefont {C.}~\bibnamefont
  {Nieto-Draghi}}, \bibinfo {author} {\bibfnamefont {J.}~\bibnamefont
  {Vrabec}}, \ and\ \bibinfo {author} {\bibfnamefont {H.}~\bibnamefont
  {Hasse}},\ }\bibfield  {title} {\enquote {\bibinfo {title} {Prediction of
  transport properties by molecular simulation: methanol and ethanol and their
  mixture},}\ }\href@noop {} {\bibfield  {journal} {\bibinfo  {journal} {J.
  Phys. Chem. B}\ }\textbf {\bibinfo {volume} {112}},\ \bibinfo {pages}
  {16664--16674} (\bibinfo {year} {2008})}\BibitemShut {NoStop}%
\bibitem [{\citenamefont {Klotz}\ and\ \citenamefont
  {Rosenberg}(2008)}]{KR2008}%
  \BibitemOpen
  \bibfield  {author} {\bibinfo {author} {\bibfnamefont {I.~M.}\ \bibnamefont
  {Klotz}}\ and\ \bibinfo {author} {\bibfnamefont {R.~M.}\ \bibnamefont
  {Rosenberg}},\ }\href@noop {} {\emph {\bibinfo {title} {Chemical
  Thermodynamics Basic Concepts and Methods}}},\ \bibinfo {edition} {7th}\ ed.\
  (\bibinfo  {publisher} {John Wiley \& Sons, Inc.},\ \bibinfo {year}
  {2008})\BibitemShut {NoStop}%
\bibitem [{\citenamefont {Staverman}(1951)}]{Staverman1951}%
  \BibitemOpen
  \bibfield  {author} {\bibinfo {author} {\bibfnamefont {A.~J.}\ \bibnamefont
  {Staverman}},\ }\bibfield  {title} {\enquote {\bibinfo {title} {The theory of
  measurement of osmotic pressure},}\ }\href@noop {} {\bibfield  {journal}
  {\bibinfo  {journal} {Recl. Trav. Chim. Pay.-B.}\ }\textbf {\bibinfo {volume}
  {70}},\ \bibinfo {pages} {344--352} (\bibinfo {year} {1951})}\BibitemShut
  {NoStop}%
\bibitem [{\citenamefont {Fritz}(1986)}]{Fritz1986}%
  \BibitemOpen
  \bibfield  {author} {\bibinfo {author} {\bibfnamefont {S.~J.}\ \bibnamefont
  {Fritz}},\ }\bibfield  {title} {\enquote {\bibinfo {title} {Ideality of clay
  membranes in osmotic processes: a review},}\ }\href@noop {} {\bibfield
  {journal} {\bibinfo  {journal} {Clay. Clay Miner.}\ }\textbf {\bibinfo
  {volume} {34}},\ \bibinfo {pages} {214--223} (\bibinfo {year}
  {1986})}\BibitemShut {NoStop}%
\bibitem [{\citenamefont {Wijmans}\ \emph {et~al.}(1985)\citenamefont
  {Wijmans}, \citenamefont {Nakao}, \citenamefont {Van Den~Berg}, \citenamefont
  {Troelstra},\ and\ \citenamefont {Smolders}}]{WNV+1985}%
  \BibitemOpen
  \bibfield  {author} {\bibinfo {author} {\bibfnamefont {J.~G.}\ \bibnamefont
  {Wijmans}}, \bibinfo {author} {\bibfnamefont {S.}~\bibnamefont {Nakao}},
  \bibinfo {author} {\bibfnamefont {J.~W.~A.}\ \bibnamefont {Van Den~Berg}},
  \bibinfo {author} {\bibfnamefont {F.~R.}\ \bibnamefont {Troelstra}}, \ and\
  \bibinfo {author} {\bibfnamefont {C.~A.}\ \bibnamefont {Smolders}},\
  }\bibfield  {title} {\enquote {\bibinfo {title} {Hydrodynamic resistance of
  concentration polarization boundary layers in ultrafiltration},}\ }\href@noop
  {} {\bibfield  {journal} {\bibinfo  {journal} {J. Membrane Sci.}\ }\textbf
  {\bibinfo {volume} {22}},\ \bibinfo {pages} {117--135} (\bibinfo {year}
  {1985})}\BibitemShut {NoStop}%
\bibitem [{\citenamefont {Gravelle}\ \emph {et~al.}(2016)\citenamefont
  {Gravelle}, \citenamefont {Ybert}, \citenamefont {Bocquet},\ and\
  \citenamefont {Joly}}]{GYB+2016}%
  \BibitemOpen
  \bibfield  {author} {\bibinfo {author} {\bibfnamefont {S.}~\bibnamefont
  {Gravelle}}, \bibinfo {author} {\bibfnamefont {C.}~\bibnamefont {Ybert}},
  \bibinfo {author} {\bibfnamefont {L.}~\bibnamefont {Bocquet}}, \ and\
  \bibinfo {author} {\bibfnamefont {L.}~\bibnamefont {Joly}},\ }\bibfield
  {title} {\enquote {\bibinfo {title} {Anomalous capillary filling and
  wettability reversal in nanochannels},}\ }\href@noop {} {\bibfield  {journal}
  {\bibinfo  {journal} {Phys. Rev. E}\ }\textbf {\bibinfo {volume} {93}},\
  \bibinfo {pages} {033123} (\bibinfo {year} {2016})}\BibitemShut {NoStop}%
\bibitem [{\citenamefont {You}\ \emph {et~al.}(2013)\citenamefont {You},
  \citenamefont {Yu}, \citenamefont {Sundqvist}, \citenamefont {Belyaeva},
  \citenamefont {Avramenko}, \citenamefont {Korobov},\ and\ \citenamefont
  {Talyzin}}]{YYS+2013}%
  \BibitemOpen
  \bibfield  {author} {\bibinfo {author} {\bibfnamefont {S.}~\bibnamefont
  {You}}, \bibinfo {author} {\bibfnamefont {J.}~\bibnamefont {Yu}}, \bibinfo
  {author} {\bibfnamefont {B.}~\bibnamefont {Sundqvist}}, \bibinfo {author}
  {\bibfnamefont {L.~A.}\ \bibnamefont {Belyaeva}}, \bibinfo {author}
  {\bibfnamefont {N.~V.}\ \bibnamefont {Avramenko}}, \bibinfo {author}
  {\bibfnamefont {M.~V.}\ \bibnamefont {Korobov}}, \ and\ \bibinfo {author}
  {\bibfnamefont {A.~V.}\ \bibnamefont {Talyzin}},\ }\bibfield  {title}
  {\enquote {\bibinfo {title} {Selective intercalation of graphite oxide by
  methanol in water/methanol mixtures},}\ }\href@noop {} {\bibfield  {journal}
  {\bibinfo  {journal} {J. Phys. Chem. C}\ }\textbf {\bibinfo {volume} {117}},\
  \bibinfo {pages} {1963--1968} (\bibinfo {year} {2013})}\BibitemShut {NoStop}%
\bibitem [{\citenamefont {Zhao}\ and\ \citenamefont {Yang}(2015)}]{ZY2015}%
  \BibitemOpen
  \bibfield  {author} {\bibinfo {author} {\bibfnamefont {M.}~\bibnamefont
  {Zhao}}\ and\ \bibinfo {author} {\bibfnamefont {X.}~\bibnamefont {Yang}},\
  }\bibfield  {title} {\enquote {\bibinfo {title} {Segregation structures and
  miscellaneous diffusions for ethanol/water mixtures in graphene-based
  nanoscale pores},}\ }\href@noop {} {\bibfield  {journal} {\bibinfo  {journal}
  {J. Phys. Chem. C}\ }\textbf {\bibinfo {volume} {119}},\ \bibinfo {pages}
  {21664--21673} (\bibinfo {year} {2015})}\BibitemShut {NoStop}%
\bibitem [{\citenamefont {Jorgensen}, \citenamefont {Maxwell},\ and\
  \citenamefont {Tirado-Rives}(1996)}]{JMT1996}%
  \BibitemOpen
  \bibfield  {author} {\bibinfo {author} {\bibfnamefont {W.~L.}\ \bibnamefont
  {Jorgensen}}, \bibinfo {author} {\bibfnamefont {D.~S.}\ \bibnamefont
  {Maxwell}}, \ and\ \bibinfo {author} {\bibfnamefont {J.}~\bibnamefont
  {Tirado-Rives}},\ }\bibfield  {title} {\enquote {\bibinfo {title}
  {Development and testing of the {OPLS} all-atom force field on conformational
  energetics and properties of organic liquids},}\ }\href@noop {} {\bibfield
  {journal} {\bibinfo  {journal} {J. Am. Chem. Soc.}\ }\textbf {\bibinfo
  {volume} {118}},\ \bibinfo {pages} {11225--11236} (\bibinfo {year}
  {1996})}\BibitemShut {NoStop}%
\bibitem [{\citenamefont {Berendsen}, \citenamefont {Grigera},\ and\
  \citenamefont {Straatsma}(1987)}]{BGS1987}%
  \BibitemOpen
  \bibfield  {author} {\bibinfo {author} {\bibfnamefont {H.~J.~C.}\
  \bibnamefont {Berendsen}}, \bibinfo {author} {\bibfnamefont {J.~R.}\
  \bibnamefont {Grigera}}, \ and\ \bibinfo {author} {\bibfnamefont {T.~P.}\
  \bibnamefont {Straatsma}},\ }\bibfield  {title} {\enquote {\bibinfo {title}
  {The missing term in effective pair potentials},}\ }\href@noop {} {\bibfield
  {journal} {\bibinfo  {journal} {J. Phys. Chem.}\ }\textbf {\bibinfo {volume}
  {91}},\ \bibinfo {pages} {6269--6271} (\bibinfo {year} {1987})}\BibitemShut
  {NoStop}%
\bibitem [{\citenamefont {Striolo}\ \emph {et~al.}(2005)\citenamefont
  {Striolo}, \citenamefont {Naicker}, \citenamefont {Chialvo}, \citenamefont
  {Cummings},\ and\ \citenamefont {Gubbins}}]{SNC+2005}%
  \BibitemOpen
  \bibfield  {author} {\bibinfo {author} {\bibfnamefont {A.}~\bibnamefont
  {Striolo}}, \bibinfo {author} {\bibfnamefont {P.~K.}\ \bibnamefont
  {Naicker}}, \bibinfo {author} {\bibfnamefont {A.~A.}\ \bibnamefont
  {Chialvo}}, \bibinfo {author} {\bibfnamefont {P.~T.}\ \bibnamefont
  {Cummings}}, \ and\ \bibinfo {author} {\bibfnamefont {K.~E.}\ \bibnamefont
  {Gubbins}},\ }\bibfield  {title} {\enquote {\bibinfo {title} {Simulated water
  adsorption isotherms in hydrophilic and hydrophobic cylindrical nanopores},}\
  }\href@noop {} {\bibfield  {journal} {\bibinfo  {journal} {Adsorption}\
  }\textbf {\bibinfo {volume} {11}},\ \bibinfo {pages} {397--401} (\bibinfo
  {year} {2005})}\BibitemShut {NoStop}%
\bibitem [{\citenamefont {Jiang}, \citenamefont {Cooper},\ and\ \citenamefont
  {Dai}(2009)}]{JCD2009}%
  \BibitemOpen
  \bibfield  {author} {\bibinfo {author} {\bibfnamefont {D.-E.}\ \bibnamefont
  {Jiang}}, \bibinfo {author} {\bibfnamefont {V.~R.}\ \bibnamefont {Cooper}}, \
  and\ \bibinfo {author} {\bibfnamefont {S.}~\bibnamefont {Dai}},\ }\bibfield
  {title} {\enquote {\bibinfo {title} {Porous graphene as the ultimate membrane
  for gas separation},}\ }\href@noop {} {\bibfield  {journal} {\bibinfo
  {journal} {Nano Lett.}\ }\textbf {\bibinfo {volume} {9}},\ \bibinfo {pages}
  {4019--4024} (\bibinfo {year} {2009})}\BibitemShut {NoStop}%
\bibitem [{\citenamefont {Nair}\ \emph {et~al.}(2012)\citenamefont {Nair},
  \citenamefont {Wu}, \citenamefont {Jayaram}, \citenamefont {Grigorieva},\
  and\ \citenamefont {Geim}}]{NWJ+2012}%
  \BibitemOpen
  \bibfield  {author} {\bibinfo {author} {\bibfnamefont {R.~R.}\ \bibnamefont
  {Nair}}, \bibinfo {author} {\bibfnamefont {H.~A.}\ \bibnamefont {Wu}},
  \bibinfo {author} {\bibfnamefont {P.~N.}\ \bibnamefont {Jayaram}}, \bibinfo
  {author} {\bibfnamefont {I.~V.}\ \bibnamefont {Grigorieva}}, \ and\ \bibinfo
  {author} {\bibfnamefont {A.~K.}\ \bibnamefont {Geim}},\ }\bibfield  {title}
  {\enquote {\bibinfo {title} {Unimpeded permeation of water through
  helium-leak--tight graphene-based membranes},}\ }\href@noop {} {\bibfield
  {journal} {\bibinfo  {journal} {Science}\ }\textbf {\bibinfo {volume}
  {335}},\ \bibinfo {pages} {442--444} (\bibinfo {year} {2012})}\BibitemShut
  {NoStop}%
\bibitem [{\citenamefont {Xia}\ \emph {et~al.}(2015)\citenamefont {Xia},
  \citenamefont {Ni}, \citenamefont {Zhu}, \citenamefont {Zhao},\ and\
  \citenamefont {Li}}]{XNZ+2015}%
  \BibitemOpen
  \bibfield  {author} {\bibinfo {author} {\bibfnamefont {S.}~\bibnamefont
  {Xia}}, \bibinfo {author} {\bibfnamefont {M.}~\bibnamefont {Ni}}, \bibinfo
  {author} {\bibfnamefont {T.}~\bibnamefont {Zhu}}, \bibinfo {author}
  {\bibfnamefont {Y.}~\bibnamefont {Zhao}}, \ and\ \bibinfo {author}
  {\bibfnamefont {N.}~\bibnamefont {Li}},\ }\bibfield  {title} {\enquote
  {\bibinfo {title} {Ultrathin graphene oxide nanosheet membranes with various
  d-spacing assembled using the pressure-assisted filtration method for
  removing natural organic matter},}\ }\href@noop {} {\bibfield  {journal}
  {\bibinfo  {journal} {Desalination}\ }\textbf {\bibinfo {volume} {371}},\
  \bibinfo {pages} {78--87} (\bibinfo {year} {2015})}\BibitemShut {NoStop}%
\bibitem [{\citenamefont {Hu}\ and\ \citenamefont {Mi}(2013)}]{HM2013}%
  \BibitemOpen
  \bibfield  {author} {\bibinfo {author} {\bibfnamefont {M.}~\bibnamefont
  {Hu}}\ and\ \bibinfo {author} {\bibfnamefont {B.}~\bibnamefont {Mi}},\
  }\bibfield  {title} {\enquote {\bibinfo {title} {Enabling graphene oxide
  nanosheets as water separation membranes},}\ }\href@noop {} {\bibfield
  {journal} {\bibinfo  {journal} {Environ. Sci. Technol.}\ }\textbf {\bibinfo
  {volume} {47}},\ \bibinfo {pages} {3715--3723} (\bibinfo {year}
  {2013})}\BibitemShut {NoStop}%
\bibitem [{\citenamefont {Ban}\ \emph {et~al.}(2016)\citenamefont {Ban},
  \citenamefont {Xie}, \citenamefont {Wang}, \citenamefont {Jing},
  \citenamefont {Liu},\ and\ \citenamefont {Zhou}}]{BXW+2016}%
  \BibitemOpen
  \bibfield  {author} {\bibinfo {author} {\bibfnamefont {S.}~\bibnamefont
  {Ban}}, \bibinfo {author} {\bibfnamefont {J.}~\bibnamefont {Xie}}, \bibinfo
  {author} {\bibfnamefont {Y.}~\bibnamefont {Wang}}, \bibinfo {author}
  {\bibfnamefont {B.}~\bibnamefont {Jing}}, \bibinfo {author} {\bibfnamefont
  {B.}~\bibnamefont {Liu}}, \ and\ \bibinfo {author} {\bibfnamefont
  {H.}~\bibnamefont {Zhou}},\ }\bibfield  {title} {\enquote {\bibinfo {title}
  {Insight into the nanoscale mechanism of rapid {H}$_2${O} transport within
  graphene oxide membrane: the impact of oxygen functional group clustering},}\
  }\href@noop {} {\bibfield  {journal} {\bibinfo  {journal} {ACS Appl. Mater.
  Interfaces}\ }\textbf {\bibinfo {volume} {8}},\ \bibinfo {pages} {321--332}
  (\bibinfo {year} {2016})}\BibitemShut {NoStop}%
\bibitem [{\citenamefont {Akbari}\ \emph {et~al.}(2016)\citenamefont {Akbari},
  \citenamefont {Sheath}, \citenamefont {Martin}, \citenamefont {Shinde},
  \citenamefont {Shaibani}, \citenamefont {Banerjee}, \citenamefont {Tkacz},
  \citenamefont {Bhattacharyya},\ and\ \citenamefont {Majumder}}]{ASM+2016}%
  \BibitemOpen
  \bibfield  {author} {\bibinfo {author} {\bibfnamefont {A.}~\bibnamefont
  {Akbari}}, \bibinfo {author} {\bibfnamefont {P.}~\bibnamefont {Sheath}},
  \bibinfo {author} {\bibfnamefont {S.~T.}\ \bibnamefont {Martin}}, \bibinfo
  {author} {\bibfnamefont {D.~B.}\ \bibnamefont {Shinde}}, \bibinfo {author}
  {\bibfnamefont {M.}~\bibnamefont {Shaibani}}, \bibinfo {author}
  {\bibfnamefont {P.~C.}\ \bibnamefont {Banerjee}}, \bibinfo {author}
  {\bibfnamefont {R.}~\bibnamefont {Tkacz}}, \bibinfo {author} {\bibfnamefont
  {D.}~\bibnamefont {Bhattacharyya}}, \ and\ \bibinfo {author} {\bibfnamefont
  {M.}~\bibnamefont {Majumder}},\ }\bibfield  {title} {\enquote {\bibinfo
  {title} {Large-area graphene-based nanofiltration membranes by shear
  alignment of discotic nematic liquid crystals of graphene oxide},}\
  }\href@noop {} {\bibfield  {journal} {\bibinfo  {journal} {Nature Commun.}\
  }\textbf {\bibinfo {volume} {7}},\ \bibinfo {pages} {10891} (\bibinfo {year}
  {2016})}\BibitemShut {NoStop}%
\bibitem [{\citenamefont {Yang}\ \emph {et~al.}(2013)\citenamefont {Yang},
  \citenamefont {Cheng}, \citenamefont {Wang}, \citenamefont {Qiu},\ and\
  \citenamefont {Li}}]{YCW+2013}%
  \BibitemOpen
  \bibfield  {author} {\bibinfo {author} {\bibfnamefont {X.}~\bibnamefont
  {Yang}}, \bibinfo {author} {\bibfnamefont {C.}~\bibnamefont {Cheng}},
  \bibinfo {author} {\bibfnamefont {Y.}~\bibnamefont {Wang}}, \bibinfo {author}
  {\bibfnamefont {L.}~\bibnamefont {Qiu}}, \ and\ \bibinfo {author}
  {\bibfnamefont {D.}~\bibnamefont {Li}},\ }\bibfield  {title} {\enquote
  {\bibinfo {title} {Liquid-mediated dense integration of graphene materials
  for compact capacitive energy storage},}\ }\href@noop {} {\bibfield
  {journal} {\bibinfo  {journal} {Science}\ }\textbf {\bibinfo {volume}
  {341}},\ \bibinfo {pages} {534--537} (\bibinfo {year} {2013})}\BibitemShut
  {NoStop}%
\bibitem [{\citenamefont {Werber}, \citenamefont {Deshmukh},\ and\
  \citenamefont {Elimelech}(2016)}]{WDE2016}%
  \BibitemOpen
  \bibfield  {author} {\bibinfo {author} {\bibfnamefont {J.~R.}\ \bibnamefont
  {Werber}}, \bibinfo {author} {\bibfnamefont {A.}~\bibnamefont {Deshmukh}}, \
  and\ \bibinfo {author} {\bibfnamefont {M.}~\bibnamefont {Elimelech}},\
  }\bibfield  {title} {\enquote {\bibinfo {title} {The critical need for
  increased selectivity, not increased water permeability, for desalination
  membranes},}\ }\href@noop {} {\bibfield  {journal} {\bibinfo  {journal}
  {Environ. Sci. Technol. Lett.}\ }\textbf {\bibinfo {volume} {3}},\ \bibinfo
  {pages} {112--120} (\bibinfo {year} {2016})}\BibitemShut {NoStop}%
\bibitem [{\citenamefont {Surblys}\ \emph {et~al.}(2014)\citenamefont
  {Surblys}, \citenamefont {Yamaguchi}, \citenamefont {Kuroda}, \citenamefont
  {Kagawa}, \citenamefont {Nakajima},\ and\ \citenamefont
  {Fujimura}}]{SYK+2014}%
  \BibitemOpen
  \bibfield  {author} {\bibinfo {author} {\bibfnamefont {D.}~\bibnamefont
  {Surblys}}, \bibinfo {author} {\bibfnamefont {Y.}~\bibnamefont {Yamaguchi}},
  \bibinfo {author} {\bibfnamefont {K.}~\bibnamefont {Kuroda}}, \bibinfo
  {author} {\bibfnamefont {M.}~\bibnamefont {Kagawa}}, \bibinfo {author}
  {\bibfnamefont {T.}~\bibnamefont {Nakajima}}, \ and\ \bibinfo {author}
  {\bibfnamefont {H.}~\bibnamefont {Fujimura}},\ }\bibfield  {title} {\enquote
  {\bibinfo {title} {Molecular dynamics analysis on wetting and interfacial
  properties of water-alcohol mixture droplets on a solid surface},}\
  }\href@noop {} {\bibfield  {journal} {\bibinfo  {journal} {J. Chem. Phys.}\
  }\textbf {\bibinfo {volume} {140}},\ \bibinfo {pages} {034505} (\bibinfo
  {year} {2014})}\BibitemShut {NoStop}%
\bibitem [{\citenamefont {Agrawal}\ and\ \citenamefont
  {Herron}(1997)}]{AH1997}%
  \BibitemOpen
  \bibfield  {author} {\bibinfo {author} {\bibfnamefont {R.}~\bibnamefont
  {Agrawal}}\ and\ \bibinfo {author} {\bibfnamefont {D.~M.}\ \bibnamefont
  {Herron}},\ }\bibfield  {title} {\enquote {\bibinfo {title} {Optimal
  thermodynamic feed conditions for distillation of ideal binary mixtures},}\
  }\href@noop {} {\bibfield  {journal} {\bibinfo  {journal} {AIChE J.}\
  }\textbf {\bibinfo {volume} {43}},\ \bibinfo {pages} {2984--2996} (\bibinfo
  {year} {1997})}\BibitemShut {NoStop}%
\bibitem [{\citenamefont {Prokopakis}\ and\ \citenamefont
  {Seider}(1983)}]{PS1983}%
  \BibitemOpen
  \bibfield  {author} {\bibinfo {author} {\bibfnamefont {G.~J.}\ \bibnamefont
  {Prokopakis}}\ and\ \bibinfo {author} {\bibfnamefont {W.~D.}\ \bibnamefont
  {Seider}},\ }\bibfield  {title} {\enquote {\bibinfo {title} {Dynamic
  simulation of azeotropic distillation towers},}\ }\href@noop {} {\bibfield
  {journal} {\bibinfo  {journal} {AIChE J.}\ }\textbf {\bibinfo {volume}
  {29}},\ \bibinfo {pages} {1017--1029} (\bibinfo {year} {1983})}\BibitemShut
  {NoStop}%
\bibitem [{\citenamefont {Holt}\ \emph {et~al.}(2006)\citenamefont {Holt},
  \citenamefont {Park}, \citenamefont {Wang}, \citenamefont {Stadermann},
  \citenamefont {Artyukhin}, \citenamefont {Grigoropoulos}, \citenamefont
  {Noy},\ and\ \citenamefont {Bakajin}}]{HPW+2006}%
  \BibitemOpen
  \bibfield  {author} {\bibinfo {author} {\bibfnamefont {J.~K.}\ \bibnamefont
  {Holt}}, \bibinfo {author} {\bibfnamefont {H.~G.}\ \bibnamefont {Park}},
  \bibinfo {author} {\bibfnamefont {Y.}~\bibnamefont {Wang}}, \bibinfo {author}
  {\bibfnamefont {M.}~\bibnamefont {Stadermann}}, \bibinfo {author}
  {\bibfnamefont {A.~B.}\ \bibnamefont {Artyukhin}}, \bibinfo {author}
  {\bibfnamefont {C.~P.}\ \bibnamefont {Grigoropoulos}}, \bibinfo {author}
  {\bibfnamefont {A.}~\bibnamefont {Noy}}, \ and\ \bibinfo {author}
  {\bibfnamefont {O.}~\bibnamefont {Bakajin}},\ }\bibfield  {title} {\enquote
  {\bibinfo {title} {Fast mass transport through sub-2-nanometer carbon
  nanotubes},}\ }\href@noop {} {\bibfield  {journal} {\bibinfo  {journal}
  {Science}\ }\textbf {\bibinfo {volume} {312}},\ \bibinfo {pages} {1034--1037}
  (\bibinfo {year} {2006})}\BibitemShut {NoStop}%
\bibitem [{\citenamefont {Majumder}\ \emph {et~al.}(2005)\citenamefont
  {Majumder}, \citenamefont {Chopra}, \citenamefont {Andrews},\ and\
  \citenamefont {Hinds}}]{MCA+2005}%
  \BibitemOpen
  \bibfield  {author} {\bibinfo {author} {\bibfnamefont {M.}~\bibnamefont
  {Majumder}}, \bibinfo {author} {\bibfnamefont {N.}~\bibnamefont {Chopra}},
  \bibinfo {author} {\bibfnamefont {R.}~\bibnamefont {Andrews}}, \ and\
  \bibinfo {author} {\bibfnamefont {B.~J.}\ \bibnamefont {Hinds}},\ }\bibfield
  {title} {\enquote {\bibinfo {title} {Enhanced flow in carbon nanotubes},}\
  }\href@noop {} {\bibfield  {journal} {\bibinfo  {journal} {Nature}\ }\textbf
  {\bibinfo {volume} {438}},\ \bibinfo {pages} {44--44} (\bibinfo {year}
  {2005})}\BibitemShut {NoStop}%
\bibitem [{\citenamefont {Booth}\ \emph {et~al.}(2008)\citenamefont {Booth},
  \citenamefont {Blake}, \citenamefont {Nair}, \citenamefont {Jiang},
  \citenamefont {Hill}, \citenamefont {Bangert}, \citenamefont {Bleloch},
  \citenamefont {Gass}, \citenamefont {Novoselov}, \citenamefont {Katsnelson},\
  and\ \citenamefont {Geim}}]{BBN+2008}%
  \BibitemOpen
  \bibfield  {author} {\bibinfo {author} {\bibfnamefont {T.~J.}\ \bibnamefont
  {Booth}}, \bibinfo {author} {\bibfnamefont {P.}~\bibnamefont {Blake}},
  \bibinfo {author} {\bibfnamefont {R.~R.}\ \bibnamefont {Nair}}, \bibinfo
  {author} {\bibfnamefont {D.}~\bibnamefont {Jiang}}, \bibinfo {author}
  {\bibfnamefont {E.~W.}\ \bibnamefont {Hill}}, \bibinfo {author}
  {\bibfnamefont {U.}~\bibnamefont {Bangert}}, \bibinfo {author} {\bibfnamefont
  {A.}~\bibnamefont {Bleloch}}, \bibinfo {author} {\bibfnamefont
  {M.}~\bibnamefont {Gass}}, \bibinfo {author} {\bibfnamefont {K.~S.}\
  \bibnamefont {Novoselov}}, \bibinfo {author} {\bibfnamefont {M.~I.}\
  \bibnamefont {Katsnelson}}, \ and\ \bibinfo {author} {\bibfnamefont {A.~K.}\
  \bibnamefont {Geim}},\ }\bibfield  {title} {\enquote {\bibinfo {title}
  {Macroscopic graphene membranes and their extraordinary stiffness},}\
  }\href@noop {} {\bibfield  {journal} {\bibinfo  {journal} {Nano Lett.}\
  }\textbf {\bibinfo {volume} {8}},\ \bibinfo {pages} {2442--2446} (\bibinfo
  {year} {2008})}\BibitemShut {NoStop}%
\bibitem [{\citenamefont {Dikin}\ \emph {et~al.}(2007)\citenamefont {Dikin},
  \citenamefont {Stankovich}, \citenamefont {Zimney}, \citenamefont {Piner},
  \citenamefont {Dommett}, \citenamefont {Evmenenko}, \citenamefont {Nguyen},\
  and\ \citenamefont {Ruoff}}]{DSZ+2007}%
  \BibitemOpen
  \bibfield  {author} {\bibinfo {author} {\bibfnamefont {D.~A.}\ \bibnamefont
  {Dikin}}, \bibinfo {author} {\bibfnamefont {S.}~\bibnamefont {Stankovich}},
  \bibinfo {author} {\bibfnamefont {E.~J.}\ \bibnamefont {Zimney}}, \bibinfo
  {author} {\bibfnamefont {R.~D.}\ \bibnamefont {Piner}}, \bibinfo {author}
  {\bibfnamefont {G.~H.~B.}\ \bibnamefont {Dommett}}, \bibinfo {author}
  {\bibfnamefont {G.}~\bibnamefont {Evmenenko}}, \bibinfo {author}
  {\bibfnamefont {S.~T.}\ \bibnamefont {Nguyen}}, \ and\ \bibinfo {author}
  {\bibfnamefont {R.~S.}\ \bibnamefont {Ruoff}},\ }\bibfield  {title} {\enquote
  {\bibinfo {title} {Preparation and characterization of graphene oxide
  paper},}\ }\href@noop {} {\bibfield  {journal} {\bibinfo  {journal} {Nature}\
  }\textbf {\bibinfo {volume} {448}},\ \bibinfo {pages} {457--460} (\bibinfo
  {year} {2007})}\BibitemShut {NoStop}%
\bibitem [{\citenamefont {Li}\ \emph {et~al.}(2013)\citenamefont {Li},
  \citenamefont {Song}, \citenamefont {Zhang}, \citenamefont {Huang},
  \citenamefont {Li}, \citenamefont {Mao}, \citenamefont {Ploehn},
  \citenamefont {Bao},\ and\ \citenamefont {Yu}}]{LSZ+2013}%
  \BibitemOpen
  \bibfield  {author} {\bibinfo {author} {\bibfnamefont {H.}~\bibnamefont
  {Li}}, \bibinfo {author} {\bibfnamefont {Z.}~\bibnamefont {Song}}, \bibinfo
  {author} {\bibfnamefont {X.}~\bibnamefont {Zhang}}, \bibinfo {author}
  {\bibfnamefont {Y.}~\bibnamefont {Huang}}, \bibinfo {author} {\bibfnamefont
  {S.}~\bibnamefont {Li}}, \bibinfo {author} {\bibfnamefont {Y.}~\bibnamefont
  {Mao}}, \bibinfo {author} {\bibfnamefont {H.~J.}\ \bibnamefont {Ploehn}},
  \bibinfo {author} {\bibfnamefont {Y.}~\bibnamefont {Bao}}, \ and\ \bibinfo
  {author} {\bibfnamefont {M.}~\bibnamefont {Yu}},\ }\bibfield  {title}
  {\enquote {\bibinfo {title} {Ultrathin, molecular-sieving graphene oxide
  membranes for selective hydrogen separation},}\ }\href@noop {} {\bibfield
  {journal} {\bibinfo  {journal} {Science}\ }\textbf {\bibinfo {volume}
  {342}},\ \bibinfo {pages} {95--98} (\bibinfo {year} {2013})}\BibitemShut
  {NoStop}%
\bibitem [{\citenamefont {Kim}\ \emph {et~al.}(2013)\citenamefont {Kim},
  \citenamefont {Yoon}, \citenamefont {Yoon}, \citenamefont {Yoo},
  \citenamefont {Ahn}, \citenamefont {Cho}, \citenamefont {Shin}, \citenamefont
  {Yang}, \citenamefont {Paik}, \citenamefont {Kwon}, \citenamefont {Choi},\
  and\ \citenamefont {Park}}]{KYY+2013}%
  \BibitemOpen
  \bibfield  {author} {\bibinfo {author} {\bibfnamefont {H.~W.}\ \bibnamefont
  {Kim}}, \bibinfo {author} {\bibfnamefont {H.~W.}\ \bibnamefont {Yoon}},
  \bibinfo {author} {\bibfnamefont {S.-M.}\ \bibnamefont {Yoon}}, \bibinfo
  {author} {\bibfnamefont {B.~M.}\ \bibnamefont {Yoo}}, \bibinfo {author}
  {\bibfnamefont {B.~K.}\ \bibnamefont {Ahn}}, \bibinfo {author} {\bibfnamefont
  {Y.~H.}\ \bibnamefont {Cho}}, \bibinfo {author} {\bibfnamefont {H.~J.}\
  \bibnamefont {Shin}}, \bibinfo {author} {\bibfnamefont {H.}~\bibnamefont
  {Yang}}, \bibinfo {author} {\bibfnamefont {U.}~\bibnamefont {Paik}}, \bibinfo
  {author} {\bibfnamefont {S.}~\bibnamefont {Kwon}}, \bibinfo {author}
  {\bibfnamefont {J.-Y.}\ \bibnamefont {Choi}}, \ and\ \bibinfo {author}
  {\bibfnamefont {H.~B.}\ \bibnamefont {Park}},\ }\bibfield  {title} {\enquote
  {\bibinfo {title} {Selective gas transport through few-layered graphene and
  graphene oxide membranes},}\ }\href@noop {} {\bibfield  {journal} {\bibinfo
  {journal} {Science}\ }\textbf {\bibinfo {volume} {342}},\ \bibinfo {pages}
  {91--95} (\bibinfo {year} {2013})}\BibitemShut {NoStop}%
\end{thebibliography}

%

\end{document}